\documentclass[aps,pre,twocolumn,superscriptaddress,showpacs]{revtex4-1}
\usepackage{graphicx}
\usepackage{amssymb}
\usepackage{epsfig}
\usepackage{amsmath}
\usepackage{times}
\usepackage{bm}

\begin{document}

\title{Growth, collapse, and self-organized criticality in complex networks}

\author{Yafeng Wang}
\affiliation{School of Physics and Information Technology, Shaanxi Normal University, Xi'an 710062, China}

\author{Huawei Fan}
\affiliation{School of Physics and Information Technology, Shaanxi Normal University, Xi'an 710062, China}

\author{Weijie Lin}
\affiliation{School of Physics and Information Technology, Shaanxi Normal University, Xi'an 710062, China}
\affiliation{Department of Physics, Zhejiang University, Hangzhou 310027, China}

\author{Ying-Cheng Lai}
\affiliation{School of Electrical, Computer, and Energy Engineering, \\ Arizona State University, Tempe, AZ 85287, USA}

\author{Xingang Wang} \email[Email address:]{wangxg@snnu.edu.cn}
\affiliation{School of Physics and Information Technology, Shaanxi Normal University, Xi'an 710062, China}

\begin{abstract}

Network growth is ubiquitous in nature (e.g., biological networks) and
technological systems (e.g., modern infrastructures). To understand how
certain dynamical behaviors can or cannot persist as the underlying
network grows is a problem of increasing importance in complex dynamical
systems as well as sustainability science and engineering. We address the
question of whether a complex network of nonlinear oscillators can maintain
its synchronization stability as it expands or grows. A network in the real
world can never be completely synchronized due to noise and/or external
disturbances. This is especially the case when, mathematically, the
transient synchronous state during the growth process becomes marginally
stable, as a local perturbation can trigger a rapid deviation of the
system from the vicinity of the synchronous state. In terms of the
nodal dynamics, a large scale avalanche over the entire network can be
triggered in the sense that the individual nodal dynamics diverge from
the synchronous state in a cascading manner within a short time period.
Because of the high dimensionality of the networked system, the transient
process for the system to recover to the synchronous state can be extremely
long. Introducing a tolerance threshold to identify the
desynchronized nodes, we find that, after an initial stage of linear
growth, the network typically evolves into a critical state
where the addition of a single new node can cause a group of nodes to
lose synchronization, leading to synchronization collapse for the entire
network. A statistical analysis indicates that, the distribution of the
size of the collapse is approximately algebraic (power law), regardless
of the fluctuations in the system parameters. This is indication of the
emergence of self-organized criticality. We demonstrate the generality of
the phenomenon of synchronization collapse using a variety of complex
network models, and uncover the underlying dynamical mechanism through
an eigenvector analysis.

\end{abstract}

\date{\today }
\maketitle

\section{Introduction} \label{sec:intro}

Growth is a ubiquitous phenomenon in complex systems. Consider, for example,
a modern infrastructure in a large metropolitan area. Due to the influx of
population, the essential facilities such as the electrical power grids,
the roads, water supply, and all kinds of services need to grow accordingly.
The issue of how to maintain the performance of the growing systems under
certain constraints (e.g., quality of living) becomes critically important
from the standpoint of sustainability. To develop a comprehensive theoretical
framework to understand, at a quantitative level, the fundamental dynamics
of sustainability in complex systems subject to continuous growth is
a challenging and open problem at the present. In this paper, to shed light
on how a complex network can maintain its function and how such a function
may be lost during growth, we focus on the dynamics of synchronization. In
particular, if a small network is synchronizable, as it grows in size the
synchronous state may collapse. The main purpose of the paper is to
uncover and understand the dynamical features of synchronization collapse
as the network grows. As will be explained, our main result is that the
collapse is essentially a self-organizing dynamical process towards
criticality with an algebraic scaling behavior.

From the beginning of modern network science, growth has been
recognized and treated as an intrinsic property of complex
networks~\cite{AB:2002,Newman:2003}. For example, the pioneering model
of scale free networks~\cite{BA:1999} had growth as a fundamental
ingredient to generate the algebraic degree distribution. The growth
aspect of this model is, however, somewhat simplistic as it stipulates
a monotonic increasing behavior in the network size, whereas
the growth behavior in real world networks can be highly non-monotonic. For example, in technological networks such as the electric power grid,
introducing a new node (e.g., a power station) will increase the load on
the existing nodes in the network, which can trigger a cascade of failures
when overload occurs~\cite{Watts:2002,ML:2002,HK:2002,MGP:2002,MPSVV:2003,Holme:2002,GLKK:2003,CLM:2004,HYY:2006,GC:2007,HLC:2008,SBPBH:2008,Gleeson:2008,YWLC:2009,Whitney:2010,WYL:2010,HL:2011,WLA:2011,LWLW:2012,Helbing:2013,GTK:2014}.
In this case, the addition of a new node does not increase the network size
but instead results in a network collapse~\cite{ML:2002,GTK:2014}. A
similar phenomenon was also observed in ecological networks, where the
introduction of a new species may result in the extinction of many
existing species~\cite{May:1972,PPP:2005}. In an economic crisis, the
failure of one financial institute can result in failures of many others
in a cascading manner~\cite{Financial_Crisis,WLA:2011}. To take
into account the phenomenon of non-monotonic network growth to avoid network
collapse, an earlier approach was to constrain the growth according to
certain functional requirement such as the system stability with respect
to certain performance, i.e., to impose the criterion that the system
must be stable at all times~\cite{May:1972}. It was revealed that network
growth subject to a global stability constraint can lead to a
non-monotonic network growth without collapse~\cite{PBTCC:2009}.
Constraint based on network synchronization was proposed~\cite{FW:2011},
where it was demonstrated that imposing synchronization stability can
result in a highly selective and dynamic growth process~\cite{FW:2011}
in the sense that it often takes many time steps for a new node to be
successfully ``absorbed'' into the existing network.

To be concrete, we study the growth of complex networks under the
constraint of synchronization stability. Synchronization of coupled
nonlinear oscillators has been an active area of research in nonlinear
science~\cite{Kuramoto:book,Strogatz:book,PRK:book,FY:1983,PC:1990},
and it is an important
type of collective dynamics on complex networks~\cite{ADGKMZ:2008}.
Earlier studies focused on systems of regular coupling structures, e.g.,
lattices or globally coupled networks. The discovery of the small
world~\cite{WS:1998} and scale free~\cite{BA:1999} network
topologies in realistic systems generated a great deal of interest in
studying the interplay between complex network structure and
synchronization~\cite{BP:2002,NMLH:2003,DHM:2005,BHLN:2005,ORHK:2005,ROH:2005,HPLYY:2006,BBH:2006,WHLL:2007,GMA:2007,GWLL:2008,HLG:2008a,HLG:2008b,WHGLL:2008,PSHMR:2014}.
Since the structures of many realistic networks are not static but
evolve with time~\cite{DM:2002,HN:2006}, synchronization in time-varying
complex networks was also studied~\cite{SBR:2006,PSBS:2006,KDL:2013}
to reveal the dynamical interplay between the time-dependent network
structure and synchronization~\cite{DHM:2005,GZ:2006,RHMRG:2009}.
We note that there was a line of works that addressed the effect on
synchronization of different ways that the network structure evolves
with time, such as link rewiring~\cite{KS:2007,SO:2008}, adjustment
of coupling weights~\cite{ZK:2006,LGL:2011}, change in the coupling
scheme~\cite{SKHJ:2009,LWFDL:2011}, but in these works the network size
is assumed to be fixed.

To investigate the growth of stability-constrained complex networks,
a key issue is the different time scales involved in the dynamical
evolution~\cite{PBTCC:2009,FW:2011,Butts:2009}. For network growth
constrained by synchronization, there are two key time scales:
one associated with the transient synchronization dynamics
occurred in a static network, denoted as $T_s$, and another
characterizing the speed of network growth, e.g., the time
interval between two successive nodal additions, $T_g$. The interplay
between the two time scales can result in distinct network evolution
dynamics. For example, for $T_s \gg T_g$, the stability constraint
would have little effect on the network evolution and, in an
approximate sense, the network grows as if no constraint were imposed.
However, for $T_s \ll T_g$, the network remains synchronized at
all times. In particular, since the stability is determined by the
network structure, e.g., through the eigenvalues of the coupling matrix,
the dynamics of network evolution is effectively decoupled from that
of synchronization. For $T_s\approx T_g$, complicated network evolution
dynamics can arise~\cite{Butts:2009}, where the two types of dynamical
processes, i.e., growth and synchronization, are entangled.
Depending on the instant network structure and synchronization
behavior, the addition of a new node may either increase
or decrease the network size. For example, if the new node induces a
desynchronization avalanche, a number of nodes will be removed if their
synchronization errors exceed some threshold values, resulting in a sudden
decrease of the network size and potentially a large scale collapse.

In this paper, we focus on the regime of $T_s\approx T_g$ and introduce
a tolerance threshold to determine if a node has become desynchronized.
Specifically, after each transient period of evolution, we remove all
nodes with synchronization error exceeding this threshold. During the
course of evolution, the network can collapse at random times.
Strikingly, we find that the size of the collapses follows an algebraic
scaling law, indicating that the network growth dynamics under the
synchronization constraint can be regarded as a process towards
self-organized criticality (SOC).

In Sec.~\ref{sec:model}, we describe our network growth model subject to
synchronization constraint and demonstrate the phenomenon of network
collapse. In Sec.~\ref{sec:statistics}, we analyze the dynamical and
statistical properties of the collapses. In Sec.~\ref{sec:theory}, we
use the method of eigenvector analysis to explain the numerically observed
collapse phenomenon. In Sec.~\ref{sec:alt_model}, we study continuous
time dynamics on randomly growing networks to demonstrate the generality
of the synchronization based collapse phenomenon and its SOC characteristics.
In Sec.~\ref{sec:conclusion}, we present conclusions and discuss the
implications of the main results.

\section{Model of network growth subject to synchronization constraint}
\label{sec:model}

We consider the standard scale-free growth model~\cite{BA:1999} but
impose a synchronization-based constraint for nodal removal. Specifically,
starting from a small, synchronizable core of $m_0$ coupled nonlinear
oscillators (nodes), at each time step $n_g$ of network growth, we
add a new node with random initial condition into the network. The new
node is connected to $m$ existing nodes according to the preferential
attachment probability $\Pi_i=k_i/\sum_{j}k_j$, where $i,j=1,2,\ldots,n$
are the nodal indices and $k_i$ is the degree of the $i$th node. We
then monitor the system evolution for a fixed time period ($T_g$) and
calculate the nodal synchronization error $\delta r_i$ (to be defined
below). Defining $\delta r_c$ as the tolerance threshold for nodal
desynchronization, if all nodes in the network meet the condition
$\delta r_i<\delta r_c$, the network size will be increased by one.
Otherwise, the nodes with $\delta r_i>\delta r_c$ will be removed from
the network, together with the links attached to them. For convenience,
we use the term ``collapse'' to describe the process of nodal removal
and the number of removed nodes, $\Delta n$, is the collapse size.

For simplicity, we set the nodal dynamics to be identical and adopt the
normalized coupling scheme~\cite{MZK:2005,WLL:2007}, where the dynamical
evolution of the $i$th oscillator in the network is governed by
\begin{equation} \label{eq:model}
\dot{\mathbf{x}}_i=\mathbf{F}(\mathbf{x}_i)
+\frac{\varepsilon}{k_i}\sum_{j=1}^{n}a_{ij}[\mathbf{H}(\mathbf{x}_j)
-\mathbf{H}(\mathbf{x}_i)],
\end{equation}
with $\mathbf{F}$ and $\mathbf{H}$ representing, respectively, the dynamics
of the isolated oscillator and the coupling function. The network structure
is characterized by the adjacency matrix $\{a_{ij}\}$, where $a_{ij}=1$ if
oscillators $i$ and $j$ are directly connected, and $a_{ij}=0$ otherwise.
The parameter $\varepsilon >0$ is the uniform coupling strength. Note
that the coupling strength from node $j$ to node $i$,
$c_{ij}=(\varepsilon a_{ij})/k_i$, in general is different from that
for the opposite direction, so the network is weighted and
directed~\cite{WLL:2007}. The class of models of linearly coupled nonlinear
oscillators with variants are commonly used in the literature of network
synchronization~\cite{PC:2015}. While Eq.~\eqref{eq:model} is for
continuous-time dynamical systems, networks of coupled nonlinear maps
can be formulated in a similar way.

\begin{figure*}[tbp]
\begin{center}
\includegraphics[width=0.9\linewidth]{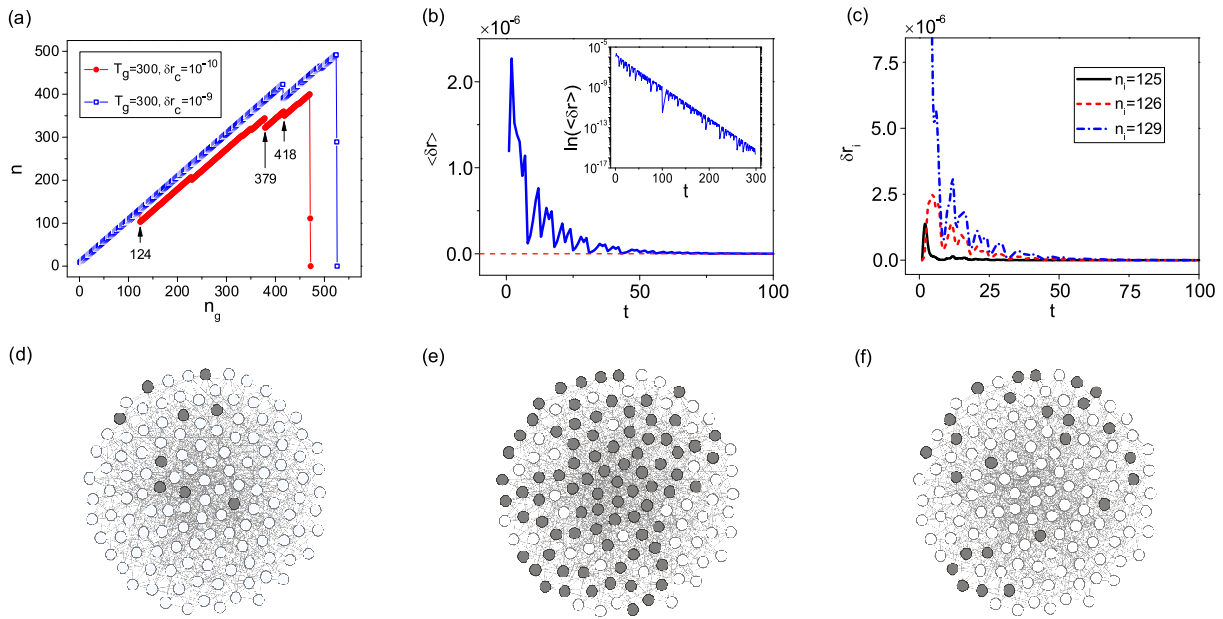}
\caption{(Color online) Evolution of a network of coupled chaotic
logistic maps subject to synchronization constraint. The transient period
for network to be synchronized is $T_g=300$, and the tolerance threshold
for desynchronization at the nodal level is $\delta r_c= 10^{-10}$.
(a) Variation of the network size, $n$, with the time step of node
addition, $n_g$. The (red) filled circles are the results for $T_g=300$
and $\delta r_c= 10^{-10}$, and the (blue) open squares are for
$T_g=300$ and $\delta r_c= 10^{-9}$. (b) Time evolution of the
network averaged synchronization error, $\left<\delta r\right>$. Inset:
the corresponding semi-logarithmic plot. (c) Time evolution of the
synchronization error, $\delta r_i$, for three typical nodes in the
network. (d-f) Snapshots of the nodal synchronization errors, $\delta r_i$,
for three different time instants: (d) $t=123T_g+1$, (e) $t=123T_g+5$,
and (f) $t=124T_g$. Nodes with $\delta r>\delta r_c$ are represented by
filled circles.}
\label{fig:growth}
\end{center}
\end{figure*}

To be concrete, we assume that the individual nodal dynamical process
is described by the chaotic logistic map, $x(t+1)=F[x(t)]=4x(t)[1-x(t)]$,
and choose $H(x)=F(x)$ as the coupling function. The coupling
strength is fixed at $\varepsilon=1$. The initial network consists of
$m_0=8$ globally coupled nodes, which is synchronizable for the given
coupling strength. For a fixed time interval $T_g=300$, we introduce
a new node (map) into the network with a randomly chosen initial condition
in the interval $(0,1)$ by attaching it to the existing nodes according
to the preferential attachment rule. The synchronization error is defined
as $\delta r_i=|x_i-\left<x\right>|$ with $\left<x\right>=\sum_i x_i/n$
being the network-averaged state, which is calculated at the end of each
time interval $T_g$. We set the tolerance threshold to be
$\delta r_c= 10^{-10}$ (somewhat arbitrarily). The growing process
is terminated either if the network has completely collapsed
($n\approx 0$) or when its size reaches a preset upper bound (e.g, 1000).

Figure~\ref{fig:growth} shows the network size $n$
versus the time step $n_g$. We see that, after an initial
period of linear growth ($n_g \le 123$), the network size is suddenly
decreased from $n=128$ to $103$, signifying that a collapse event of size
$\Delta n=25$ has occurred after the addition of the $124$th oscillator.
After the collapse, the network begins to expand again. In the subsequent
time evolution, collapse of different sizes occurs at random times, e.g.,
$\Delta n=22$ at $n_g=379$ and $\Delta n=10$ at $n_g=418$. For relatively
small network size, when a collapse event occurs, the removed nodes
account for only a small fraction of the nodes in the entire network
(e.g., $\Delta n/n < 10\%$), with growth followed immediately after
the collapse. However, as the network size exceeds a critical value,
say $n_{max}=400$, this scenario of small-scale collapse followed by
growth is changed dramatically. As shown in Fig.~\ref{fig:growth}(a),
for $n_g=471$, a catastrophic collapse event occurs, which removes over
$75\%$ of the nodes in the network (from $471$ to $111$). More strikingly,
there is no growth after the event - the network continues to collapse.
At the end of $n_g = 472$, not a single node remains in the network,
i.e., the network has collapsed {\em completely}.

To gain more insights into the dynamics of network collapse, we monitor
the system evolution for the time period $123T_g<t<124T_g$, i.e., the
response of the network dynamics to the addition of the $124$th node.
Figure~\ref{fig:growth}(b) shows the time evolution of the averaged
network synchronization error,
$\left<\delta r\right>=\sum_i \delta r_i/n$, where its value approaches
zero rapid with time. A semi-logarithmic plot reveals an exponentially
decreasing behavior for $\left<\delta r\right>$ [inset of
Fig.~\ref{fig:growth}(b)], indicating that the network is able to restore
synchronization for relatively large values of $T_g$. However, for $T_g=300$,
at the end of the time interval $t=124T_g$, the synchronization errors of
certain nodes exceed the threshold, leading to their removal from
the network. The synchronization errors for three typical nodes are
shown in Fig.~\ref{fig:growth}(c). Examining the individual nodal
synchronization errors $\delta r_i$, we find that, the ``disturbance''
triggered by the addition of a new node spreads quickly over the network,
as shown in Fig.~\ref{fig:growth}(d). After the disturbance reaches the
maximal dynamical range at $t\approx 123T_g+5$ [Fig.~\ref{fig:growth}(e)], it
begins to shrink and, at the end of this time interval, there are still a
few nodes with $\delta r>\delta r_c$, as shown in Fig.~\ref{fig:growth}(f).
Based on their dynamical responses, the nodes can be roughly
divided into three categories, as shown in Fig.~\ref{fig:growth}(c).
Specifically, for most nodes, as time increases $\delta r$ first increases
and then decreases, e.g., the $126$th node. There are also nodes for which
the values of $\delta r$ decrease monotonically with time, e.g., the $125$th
node. Finally, there are a few nodes for which the values of $\delta r$
remain about $0$, e.g., the $129$th node. We also observe that, sometimes,
the new node, whose introduction into the network triggers a network collapse,
in fact remains in the network.

\section{Statistical properties of collapse and self-organized
criticality} \label{sec:statistics}

In terms of practical significance, the following questions about network
collapse are of interest: (1) what kind of nodes are more likely to be
removed? (2) what is the size distribution of the collapse? (3) how
frequent is the network collapsed? and (4) what are the effects of
the tolerance threshold $\delta r_c$ and growing interval $T_g$ on the
collapse? In this section we address these questions numerically.

\begin{figure}[tbp]
\includegraphics[width=0.8\linewidth]{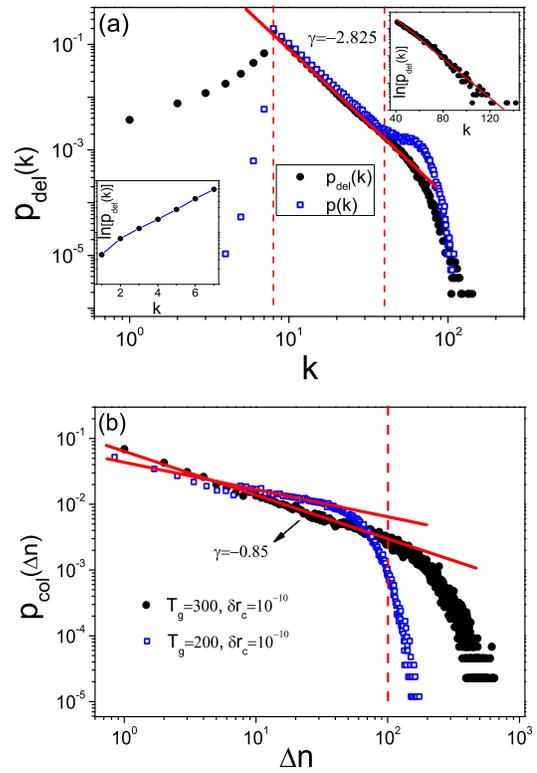}
\caption{(Color online) Statistical properties of collapse and SOC.
(a) Degree distribution $p_{del}(k)$ of the removed nodes (filled circles).
For $k\in[m,40]$, the scaling behavior is $p_{del}(k)\sim k^{\gamma}$, with
$\gamma\approx -2.83$. For $k\le m$ and $k\ge 40$, $p_{del}(k)$ increases
and decreases with $k$ exponentially, respectively. Open squares are for
the degree distribution $p(k)$ of the generated network. (b) Size
distribution $p_{col}(\Delta n)$ of the collapse event for parameters
$T_{g}=300$ and $\delta r_c= 10^{-10}$. For $\Delta n\in[1,100]$,
the scaling is $p_{col}(\Delta n)\sim \Delta n^{\gamma}$ with
$\gamma\approx -0.85$. Open squares are for the size distribution of the
collapse events for $T_{g}=200$ and $\delta r_c= 10^{-10}$. The algebraic
scaling of the collapse size signifies SOC. The results are averaged over
$100$ network realizations.}
\label{fig:statistics}
\end{figure}

A simple way to identify the removed nodes is to examine their degrees.
With the same parameters as in Fig.~\ref{fig:growth}, we plot in
Fig.~\ref{fig:statistics}(a) the normalized degree distribution, $p_{del}(k)$,
of the removed nodes collected from a large number of collapse events
(except the catastrophic one that totally destroys the network). We see
that the distribution contains approximately three distinct segments with
different scaling behaviors. Specifically, for $k\in[1,m]$, $p_{del}(k)$
increases with $k$ exponentially. For $k\in[m,40]$, $p_{del}(k)$ decreases
with $k$ algebraically with the exponent $\gamma\approx -2.83$. For
$k\in[40, 120]$, $p_{del}(k)$ decreases with $k$ exponentially. Since,
in our model each new node has $m=8$ links, it is somewhat surprising to
see from Fig.~\ref{fig:statistics}(a) that some nodes have their degrees
smaller than $m$. This phenomenon can be attributed to the node removal
mechanism: when a node is removed, all links associated to it are also
removed. Another phenomenon is that $p_{del}(k)$ reaches its maximum
at $k=8$, which seems to contradict the previous result that nodes of
large degrees are more stable with respect to synchronization than those of
small degrees~\cite{MZK:2005,ZK:2006,WLL:2007,PC:2015}.

Since $p_{del}(k)$ is obtained from a large number of collapses, to uncover
the interplay between nodal stability and degree, we need to take into
account the degree distribution $p(k)$ of the generated network. To find
$p(k)$, we use the largest network emerged in the growth process (the
network formed immediately before the catastrophic collapse) and obtain the
degree distribution for an ensemble of such networks. The results are
also shown in Fig.~\ref{fig:statistics}(a). We see that the two
distributions, $p_{del}(k)$ and $p(k)$, coincide with each other well,
where $p(k)$ also contains three distinct segments and reaches its
maximum at $k=m$. The consistency between $p_{del}(k)$ and $p(k)$
suggests that the nodal stability is independent of the degree.
Statistically, we thus expect that the small and large degree nodes
to have equal probability to be removed.

Figure~\ref{fig:statistics}(b) shows the collapse size distribution,
where the catastrophic network size $n_{max}$ is not included.
We see that, in the interval $\Delta n\in [1,100]$, the distribution
follows an algebraic scaling: $p_{col}(\Delta n)\sim \Delta n^{\gamma}$,
with $\gamma\approx -0.85$. For $\Delta n>100$, an exponential tail is
observed. To test whether the exponential tail is a result of the finite
size effect, we decrease the transient period to $T_g=200$ and plot the
distribution of the collapse size again. (As we will demonstrate later,
as $T_g$ is decreased, the maximum network size $n_{max}$ will decrease
monotonically.) Figure~\ref{fig:statistics}(b) indicates that, comparing
with the case of $T_g=300$, the regime of algebraic scaling is shifted
toward the left for $T_g=200$. Specifically, for $T_g=200$, we have
$p_{col}(\Delta n)\sim \Delta n^{\gamma}$ in the interval
$\Delta n\in [1,50]$, where the fitted exponent is about $-0.79$.

The emergence of algebraic scaling in the size distribution of network
collapse is interesting from the viewpoint of SOC that occurs in many
real-world complex systems. For a dynamical system subject to continuous
external perturbations, during its evolution towards SOC, it can appear
stable for a long period of time before a catastrophic event occurs, and
the probability for the catastrophe can be markedly larger than
intuitively expected (algebraic versus exponential
scaling)~\cite{BTW:1987,Stanley:book}. In our case, there is a long time
period of synchronization stability in spite of the small-size collapses, but
catastrophic collapses that remove all or most of the nodes in the network
can occur, albeit rarely. There are a variety of models for SOC, but the
unique feature of our model is that it exploits network synchronization
stability as a mechanism for catastrophic failures. Since synchronization
is ubiquitous in natural and man-made complex systems, the finding of SOC
in synchronization-stability-constrained network may have broad implications.
For instance, synchronization is commonly regarded as the dynamical basis
for normal functioning of the power grids~\cite{MMAN:2013}, and
there is empirical evidence that the size of the blackouts follows roughly
an algebraic distribution~\cite{PA:2013}.

\begin{figure}[tbp]
\includegraphics[width=0.8\linewidth]{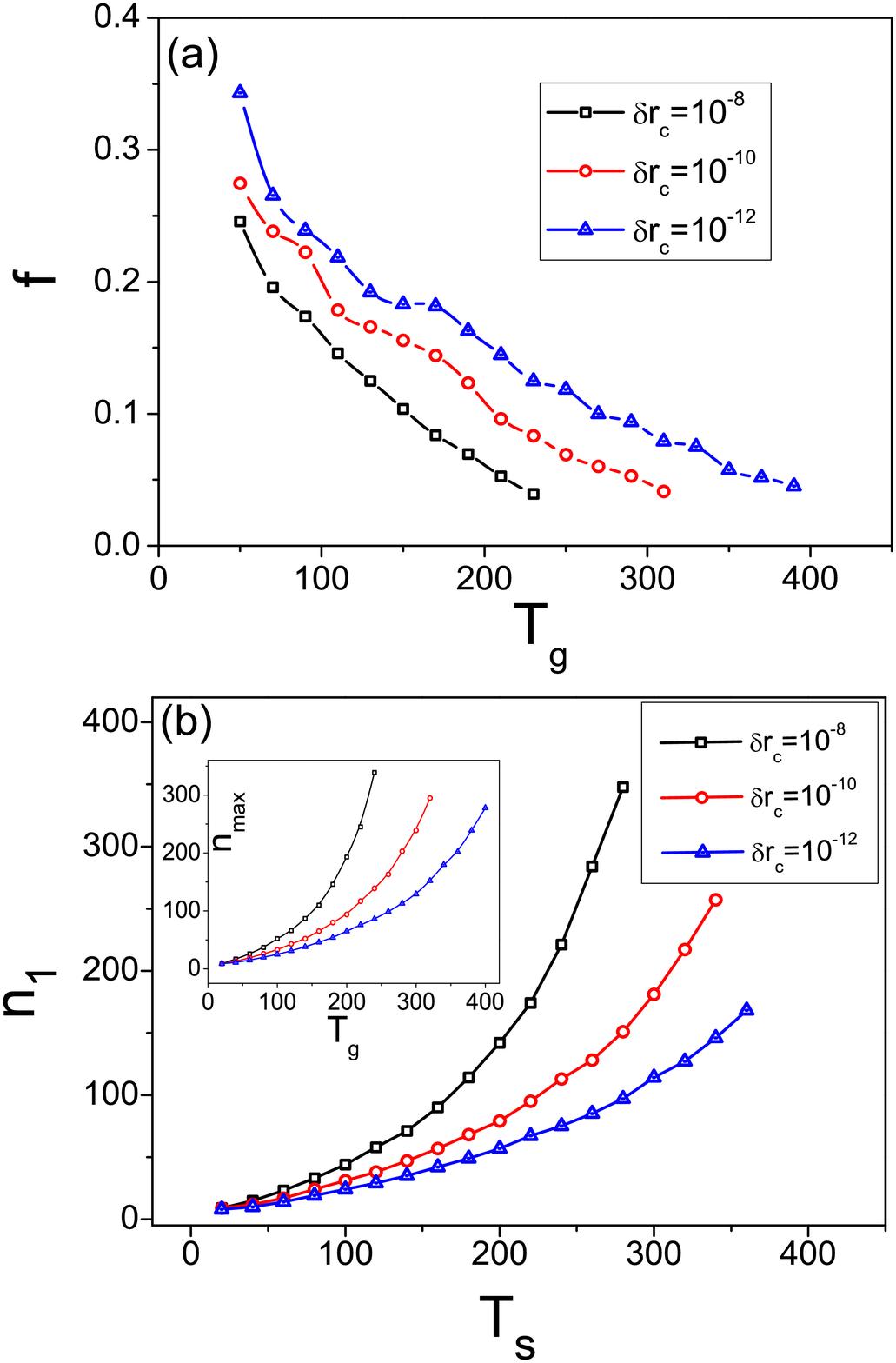}
\caption{(Color online) Behavior of the collapse frequency.
(a) The collapse frequency $f$ as a function of the transient interval $T_g$
for different values of the tolerance threshold $\delta r_c$. (b) The
first critical network size $n_1$ versus $T_g$ for different
values of $\delta r_c$. Inset: dependence of the maximum network size
$n_{max}$ on $T_g$. The results are averaged over $100$ network
realizations.}
\label{fig:frequency}
\end{figure}

We proceed to study the frequency of network collapse. Let $\Delta n'$ be
the period of continuous network growth, i.e., the number of nodes
successively added into the network between two adjacent collapses.
The collapse frequency is $f=1/\left<\Delta n'\right>$, where
$\left<\Delta n'\right>$ is the averaged period.
For the same parameters in Fig.~\ref{fig:growth}, we find
$f\approx 1/21$. That is, on average the network collapses every $21$
new additions. Since the synchronization errors are evaluated at the
end of each transient interval and nodes are removed according to a
predefined tolerance threshold, we expect the collapse
frequency to depend on the parameters $T_g$ and $\delta r_c$. This
is apparent in Fig.~\ref{fig:growth}(a), where the network growth
under the parameters $T_g=300$ and $\delta r_c= 10^{-9}$ is also
shown. We see that, comparing with the case of $\delta r_c= 10^{-10}$,
the catastrophic collapse is postponed. To assess the influence of $T_g$
and $\delta r_c$ on $f$, we show in Fig.~\ref{fig:frequency}(a) $f$
versus $T_g$ for different values of $\delta r_c$. It can be seen that,
with the increase of $T_g$ or $\delta r_c$, $f$ decreases monotonically.

For the process of network growth, two particularly relevant quantities
are: (1) the critical network size $n_1$ at which the first collapse
occurs and (2) the maximum network size $n_{max}$ beyond which a
catastrophic collapse occurs. Similar to the collapse frequency, these
two quantities depend on the parameters $T_g$ and $\delta r_c$.
Figure~\ref{fig:frequency}(b) shows $n_1$ ($n_{max}$) versus $T_g$ for
different values of $\delta r_c$. We see that, as $T_g$ or $\delta r_c$
is increased, $n_1$ ($n_{max}$) increases monotonically. That is, by
increasing $T_g$ or $\delta r_c$, one can postpone the first and the
catastrophic network collapse but eventually it will occur.

\section{Physical theory of synchronization based network collapse}
\label{sec:theory}

Say at step $n'$ of the growth, the network contains $n-1$ synchronized
oscillators and a new oscillator of random initial condition is introduced.
Due to the new oscillator, the trajectories of the existing oscillators
leave, at least temporarily, the synchronous manifold $\mathbf{x}_s$. Let
$\delta \mathbf{x}_i=\mathbf{x}_i-\mathbf{x}_s$ be the distance of the
$i$th oscillator from the manifold, which is the synchronization error.
The evolution of $\delta \mathbf{x}_i$ is governed by the following
variational equation:
\begin{equation} \label{eq:var}
\delta\dot{\mathbf{x}}_i = \mathbf{DF}(\mathbf{x}_s) \cdot \delta\mathbf{x}_i
+ \frac{\varepsilon}{k_i}\sum_{j=1}^{n}a_{ij}
\mathbf{DH}(\mathbf{x}_s)\cdot [\delta \mathbf{x}_j - \delta\mathbf{x}_i],
\end{equation}
where $\mathbf{DF(x_s)}$ and $\mathbf{DH(x_s)}$ are the Jacobian matrices
of the local dynamics and the coupling function evaluated on $\mathbf{x}_s$,
respectively. Equation~\eqref{eq:var} is obtained by linearizing
Eq.~\eqref{eq:model} about the synchronous manifold $\mathbf{x}_s$, which
characterizes its local stability~\cite{PC:1998}. To keep the expanded
network synchronizable, a necessary condition is that all the synchronization
errors, $\{\delta \mathbf{x}_i\}$ approach zero exponentially with time.
Projecting $\delta \mathbf{x}_i$ into the eigenspace spanned by the
eigenvector $\mathbf{e}_i$ of the network coupling matrix
$C=\{c_{ij}\}=\{\varepsilon a_{ij}/k_i\}$, we can diagonalize the $n$ coupled variational
equations into $n$ decoupled modes in the blocked form
\begin{equation} \label{eq:block}
\dot{\bm{\xi}}_{l}=\left[\mathbf{DF}(\mathbf{x}_s) +
\sigma \mathbf{DH}(\mathbf{x}_s \right] \cdot \bm{\xi}_{l}, l=1,\ldots,n,
\end{equation}
where $\bm{\xi}_{l}$ is the $l$th mode transverse to the synchronous
manifold $\mathbf{x}_s$, and $0=\sigma_1>\sigma_2> \ldots > \sigma_n$
are the eigenvalues of the coupling matrix $C$. Among the $n$ modes,
the one associated with $\sigma =0$ represents the motion within the
synchronous manifold. The network is synchronizable only when all the
transverse modes ($\bm{\xi}_j, j=2,\ldots,n$) are stable, i.e.,
the largest Lyapunov exponent among these modes should be negative:
$\Lambda (\sigma)<0$. For typical nonlinear oscillators and smooth
coupling functions, previous works~\cite{PC:1998,HYL:1998,HCLP:2009} showed
that $\Lambda (\sigma)$ can be negative within a bounded region in the
parameter space of $\sigma$, i.e., $\Lambda(\sigma)<0$ for
$\sigma\in\left(\sigma_{l}, \sigma_{r}\right)$. Thus, the necessary
condition to make the synchronous state stable is
$\sigma_l<\sigma_j<\sigma_r$ for all the transverse modes ($j=2,\ldots,n$).
For the chaotic logistic map used in our numerical simulations, we have
$\sigma_l=0.5$ and $\sigma_r=1.5$.

The eigenvalue analysis, also known as the master stability function (MSF)
analysis, is standard in synchronization analysis~\cite{PC:1998,HYL:1998}.
It not only indicates whether a network is synchronizable, but also
quantifies the degree of synchronization stability as well as the
synchronization speed in certain situations~\cite{Wackerbauer:2007,QHSWC:2008,
BGT:2008}. Specifically, by examining the Lyapunov exponents associated
with the two extreme modes, $\Lambda(\sigma_2)$ and $\Lambda(\sigma_n)$,
one can predict whether the network is synchronizable and how stable
(unstable) the synchronous state is. In general, the smaller
$\Lambda(\sigma_2)$ and $\Lambda(\sigma_n)$ are, the more stable the
synchronous state is~\cite{PC:1998,HYL:1998,HCLP:2009}.
Because of the relation $\Lambda(\sigma_{2,n})\propto \sigma_{2,n}$, near
the critical points $\sigma_l$ and $\sigma_r$, the network synchronizability
can be characterized by the stability distances $d_{l}=\sigma_{2}-\sigma_{l}$
and $d_{r}=\sigma_{r}-\sigma_{n}$. For a synchronizable network, we
have $d_{l,r}>0$. Moreover, the larger $d_l$ and $d_r$ are, the more stable
the synchronous state will be. Otherwise, if one of the distances is
negative, the synchronous state will be unstable. In the asynchronous
case, the smaller $d_l$ and $d_r$ are, the more unstable the synchronous
state will be.

\begin{figure}[tbp]
\includegraphics[width=0.8\linewidth]{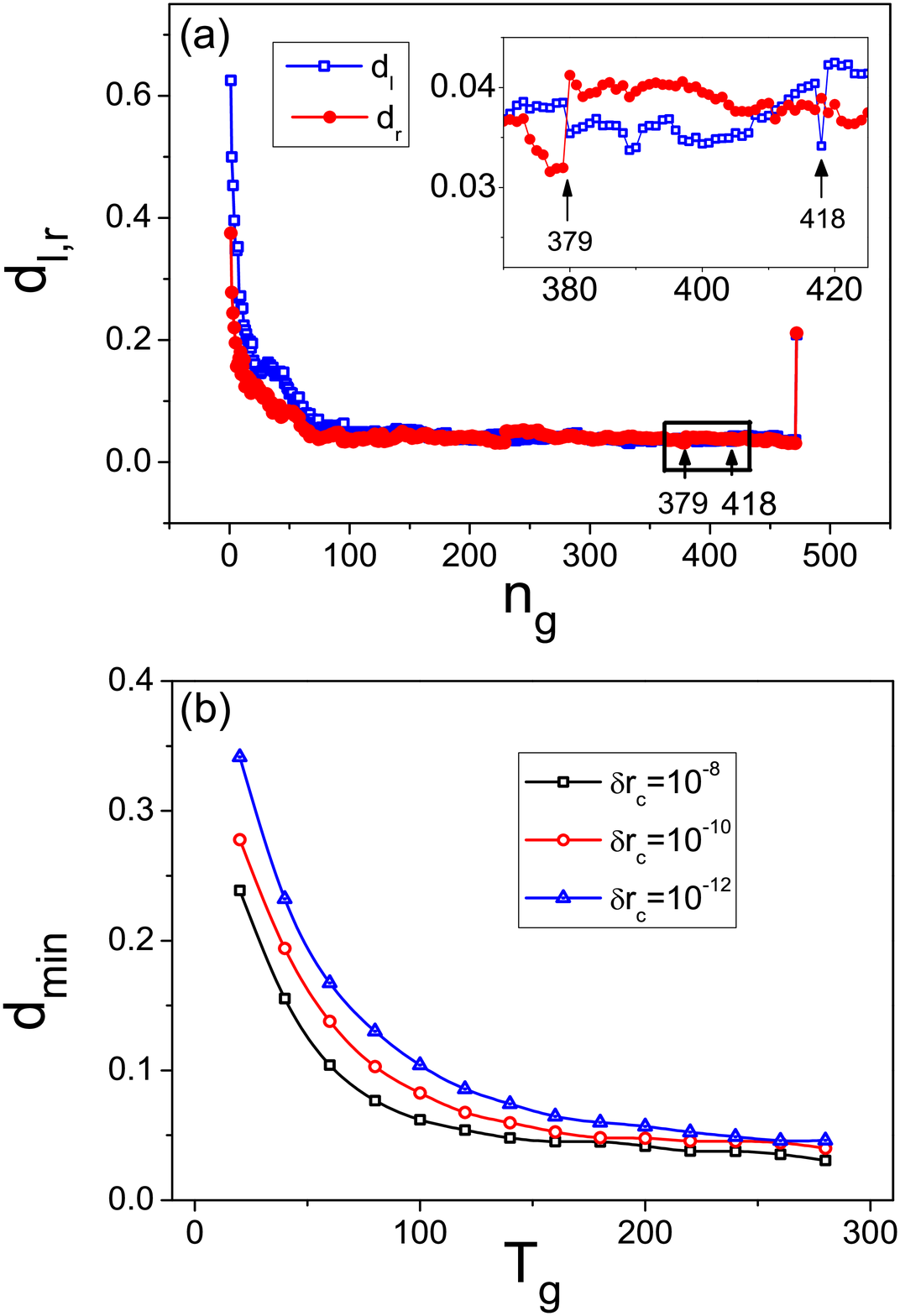}
\caption{(Color online) Behavior of synchronization distances.
(a) Time evolution of the stability distances $d_{l,r}$. Inset: a
magnification of part of the evolution. (b) The smallest stability
distance $d_{min}$ versus the transient interval $T_g$ for different
values of the tolerance threshold $\delta r_c$. The results are
averaged over $100$ network realizations.}
\label{fig:distance}
\end{figure}

As the network synchronizability can be characterized by the stability
distances $d_{l,r}$, we calculate the evolution of $d_{l,r}$ during
the course of network growth, as shown in Fig.~\ref{fig:distance}(a).
In accordance with the process of network growth (Fig.~\ref{fig:growth}),
the time evolution of $d_{l,r}$ also consists of distinct regimes. Firstly,
as $n_g$ increase from 1 to $123$, $d_{l,r}$ approaches zero quickly.
Secondly, in the interval $n_g\in(123,470)$, $d_{l,r}$ remains about zero.
A magnification of this interval reveals that, while $d_{l,r}$ tend to
reach zero, the process is occasionally interrupted by some small
increments. Checking the points at which $d_{l,r}$ increase suddenly [inset
of Fig.~\ref{fig:distance}(a)], we find that these points correspond to
exactly the time instants of network collapses. For example, for
$n_g=379$, $d_l$ increases from $0.032$ to $0.041$
[Fig.~\ref{fig:distance}(a)], while at the same time there is a collapse
event in which the network size changes from $n=344$ to $322$
[Fig.~\ref{fig:growth}(a)]. Finally, at the critical instant $n_g=472$
where the catastrophic collapse occurs, $d_{l}$ and $d_{r}$ change suddenly
to $0.21$ and $0.22$, respectively.

Figure~\ref{fig:distance} thus indicates that, for the entire process of
network growth, the stability distances $d_{l,r}$ remain positive so that
the network is synchronizable at all time. That is, even at the time when
a collapse occurs, no node would be removed if the transient time $T_g$ is
sufficiently long. It may then be said that, with respect to the impact of
the network synchronizability (as determined by the network structure),
network collapse is equally influenced by the transient synchronization
dynamics. Increasing $T_g$ can thus effectively postpone the collapses
as the network grows, a manifestation of which is a further decrease in
$d_{l,r}$ at the collapses. Let $d_{min}$ be the minimum of $d_{l,r}$
during the process of network growth. Figure~\ref{fig:distance}(b)
shows $d_{min}$ versus $T_g$ for different values of $\delta r_c$. As
anticipated,  increasing the value of $T_g$ or $\delta r_c$ results in
a monotonic decrease in the value of $d_{min}$, which agrees with the
results of direct simulations in Fig.~\ref{fig:frequency}(b) where
a postponement of the catastrophic collapse is explicitly demonstrated.

\begin{figure}[tbp]
\includegraphics[width=0.8\linewidth]{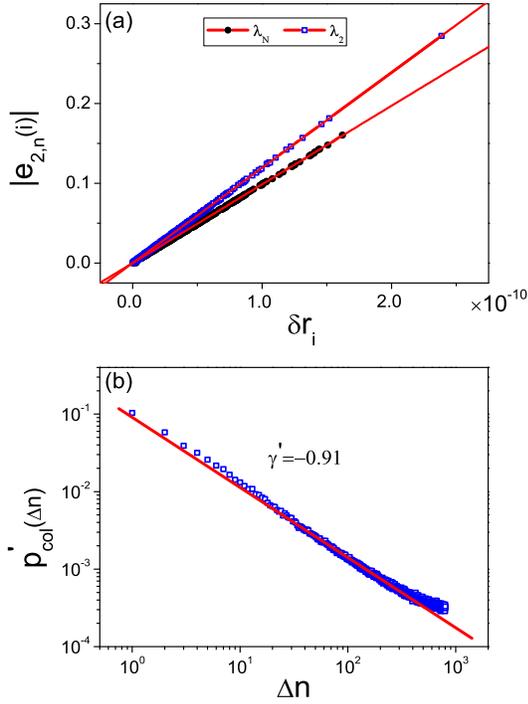}
\caption{(Color online) Relation between key eigenvector and
synchronization error. (a) The linear relationship between the absolute
eigenvector elements $|e_{2,n}(i)|$ and the oscillator synchronization errors
$\delta r_i$ at different steps of the network growth. Filled circles are
for the case of $n_g=418$, $\Lambda(\sigma_2)>\Lambda(\sigma_n)$, where
the relation $|e_{2}(i)|\sim \delta r_i$ holds. Open squares specify
the case of $n_g=379$ and $\Lambda(\sigma_2)<\Lambda(\sigma_n)$ where
we have $|e_{n}(i)|\sim \delta r_i$. (b) Size distribution of network
collapse predicted from the eigenvector analysis. The distribution follows
an algebraic scaling law: $p'_{col}(\Delta n)\sim \Delta n^{\gamma'}$,
with the fitted exponent being $\gamma'\approx -0.91$.}
\label{fig:eigenvalue}
\end{figure}

The fact that $d_{l,r}$ become approximately zero prior to a catastrophic
collapse implies that the network becomes marginally stable during the growing
process, i.e., the oscillator trajectories deviate only slightly from the
synchronous manifold. In this case, desynchronization is determined by the
two extreme modes, $\sigma_2$ and $\sigma_n$, as the corresponding transverse
Lyapunov exponents $\Lambda(\sigma_{2,n})$ are larger than those associated
with other transverse modes~\cite{FZZW:2012}. This feature makes possible
a theoretical analysis of the collapse phenomenon. In particular, assuming
$d_{l,r}\approx 0$ and $\Lambda(\sigma_{2})>\Lambda(\sigma_{n})$ (so that
the $2$nd transverse mode is more unstable), we have that desynchronization
is mainly determined by the $2$nd mode, with
$\xi_2(t)\sim \exp[\Lambda(\sigma_2)t]$. Since $\Lambda(\sigma_2)\approx 0$,
we have $\xi_2(t)\sim \Lambda(\sigma_2)t$. Transforming this mode back to
the nodal space, we obtain
$\delta r_i=|e_{2,i}\xi_2|\sim |e_{2,i}\Lambda(\sigma_2)t|$, where
$e_{2,i}$ is the $i$th component of the eigenvector $\bm{e}_2$ associated
with $\sigma_2$. For the given network structure, the value of
$\Lambda(\sigma_2)$ is fixed. We thus have
\begin{equation} \label{eq:delta_r_i}
\delta r_i\propto |e_{2,i}|,
\end{equation}
which establishes a connection between the network structure and
the oscillator stability. It is only necessary to calculate the eigenvector
associated with the most unstable mode to identify the unstable oscillators,

Relation~\eqref{eq:delta_r_i} can be verified numerically. As shown in the
inset of Fig.~\ref{fig:distance}(a), at the growing step $n_g=379$, the
network contains $n=322$ oscillators and the two extreme eigenvalues
are $(\sigma_2,\sigma_n)=(0.538,1.468)$. Since $\Lambda(\sigma_2)=-0.079$
and $\Lambda(\sigma_n)=-0.066$, desynchronization is determined by
the $n$th mode. Figure~\ref{fig:eigenvalue}(a) shows the synchronization
errors (measured at the end of the $379$th growing step) $\delta r_i$
versus the absolute eigenvector element $|e_{2,i}|$ for all the oscillators
in the network, which is obtained from the network coupling matrix $C$.
We see that $\delta r_i$ increases with $|e_{n,i}|$ linearly. The linear
relationship is also observed when the $2$nd transverse mode is more
unstable. For example, at the growing step $n_g=418$, the network contains
$n=350$ oscillators and the two pertinent Lyapunov exponents are
$[\Lambda(\sigma_2),\Lambda(\sigma_n)]=(-0.070,-0.081)$. The linear
variation of $\delta r_i$ with $|e_{2,i}|$ is also shown in
Fig.~\ref{fig:eigenvalue}(a).

Relation~\eqref{eq:delta_r_i} can also be used to interpret the size
distribution of the network collapses observed numerically [e.g.,
Fig.~\ref{fig:statistics}(b)]. Let $\delta r_i(0)$ be the initial
synchronization error of the $i$th oscillator induced by the newly
added oscillator. After a transient phase of length $T_g$, the error becomes
$\delta r_i\approx \delta r_i(0)|e_{j',i}|\exp{[\Lambda(\sigma_{j'})T_g]}$,
with $j'=2$ or $n$ (depending on which mode is more unstable). As
$\Lambda(\sigma_{j'})$ is approximately zero, we have
$\delta r_i \approx \delta r_i(0)|e_{j',i}|[1+\Lambda(\sigma_{j'})T_g]$.
Setting $\delta r_i=\delta r_c$, we get the critical element
\begin{displaymath}
e_c \approx \delta r_c/[\delta r_i(0)(1+\Lambda(\sigma_{j'})T_g)].
\end{displaymath}
Thus, whether the $i$th oscillator is removed solely depends on the
element $e_{j',i}$. In particular, if $|e_{j',i}|>e_c$, we have
$\delta r_i > \delta r_c$ so that the oscillator will be removed; otherwise
it will remain in the network. Assuming the oscillators have the same
initial error $\delta r(0)$, we can estimate the size of the network collapse
simply by counting the number of elements satisfying the inequality
$|e_{j',i}|>e_c$. To verify this idea, we generate scale-free networks,
calculate the eigenvector $\bm{e}_2$, and identify the largest element
$e_{max}$ of $\bm{e}_2$. Choosing $e_c$ randomly from the range $(0,e_{max})$
[since $d(0)$ is dependent upon the (random) initial condition of the newly
added oscillator], we truncate the eigenvector elements, where the number of
truncated elements is the collapse size. We repeat this truncation procedure
for a large number of statistical realizations and calculate the size
distribution of the collapses. The result for a network of size $n=800$ is
shown in Fig.~\ref{fig:eigenvalue}(b). We see that the size distribution
calculated from the eigenvector analysis also follows an algebraic scaling:
$p'_{col}(\Delta n)\sim \Delta n^{\gamma'}$, where the fitted exponent
is $\gamma'\approx -0.91$. This is in good agreement with the one obtained
from direct simulations [Fig.~\ref{fig:statistics}(b)], where the algebraic
scaling exponent is $\gamma\approx -0.85$ for the interval
$\Delta n\in [1,100]$.

\section{Alternative models of network dynamics} \label{sec:alt_model}

\begin{figure}[tbp]
\includegraphics[width=0.8\linewidth]{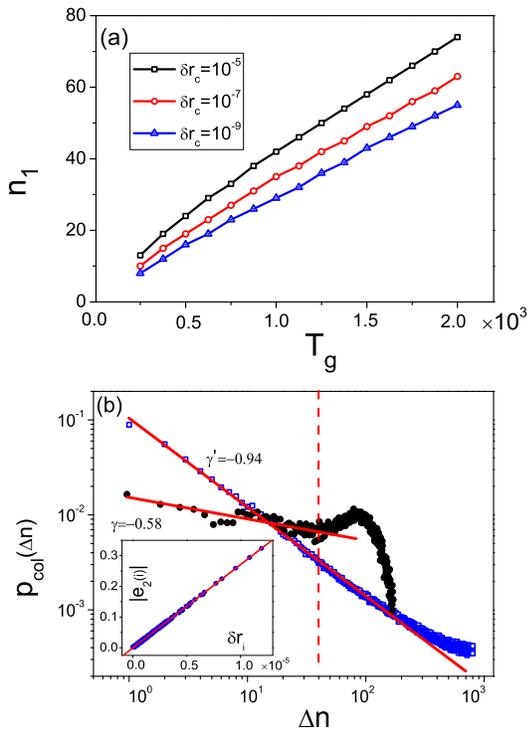}
\caption{(Color online) Synchronization based collapse in networks
of continuous-time nonlinear oscillators. For networks of chaotic
R\"{o}ssler oscillators formed according to the random link attachment
rule, the network collapse phenomenon and its SOC characteristics:
(a) the critical network size $n_1$ versus the transient time $T_g$
for different values of the tolerance threshold $\delta r_c$ and
(b) distribution of the collapse sizes
for $\Delta n\in [1,40]$: $p_{col}(\Delta n)\sim \Delta n^{\gamma}$ with
$\gamma\approx -0.58$. Open squares represent the size distribution
predicated from the eigenvector analysis. Inset: the linear relation
between $|e_{2,i}|$ and $\delta r_i$ as predicted
[Relation~\eqref{eq:delta_r_i}]. The data are averaged over $100$
network realizations.}
\label{fig:rossler}
\end{figure}

To demonstrate the generality of the synchronization based network
collapse phenomenon and its SOC characteristics, we simulate continuous
time dynamics on networks that grow according to alternative rules other
than the preferential attachment mechanism. In fact, in network modeling,
the way by which a new node is added to the existing network can have
a determining role in the network structure~\cite{AB:2002}. For example,
in unconstrained growing networks, random attachment cannot lead to
any scale free feature but results in an exponential degree
distribution~\cite{BAJ:1999}. Since the network structure has a
significant effect on synchronization, we expect the characteristics of
network growth dynamics following random attachment to be different from
those from the preferential attachment rule. Besides the network
structure, our eigenvector analysis indicates that the synchronization
behavior is also dependent upon the nodal dynamics and the coupling
function. For example, for a different type of nodal dynamics, the
MSF curve can be dramatically different, so is the stability parameter
region~\cite{PC:1998,HYL:1998,HCLP:2009}. We are led by these
considerations to study continuous-time oscillator networks that
grow according to the random attachment rule.

We choose the chaotic R\"{o}ssler oscillator~\cite{Rossler:1976} described by
$(dx/dt,dy/dt,dz/dt) = (-y-z,x+0.2y,0.2+xz-9.0z)$. The oscillators at
different nodes are coupled through the $x$ variable with the coupling
function $\mathbf{H}([x,y,z]^T ) = [0,y,0]^T$. We define the synchronization
error as $\delta r_i=|x_i-\left<x\right>|$. The coupling strength is fixed
at $\varepsilon=0.35$. The stable synchronization region from the MSF curve
is open at the right side~\cite{HCLP:2009}, i.e., the transverse mode $i$ is
stable for $\sigma_i>\sigma_l\approx 0.157$. Adopting the random attachment
rule, we grow the network under the constraint of synchronization
stability and find the phenomenon of network collapse to be robust. For
example, Fig.~\ref{fig:rossler}(a) shows the critical network size $n_1$ versus
the transient time $T_g$ for different values of the tolerance threshold
$\delta r_c$. We see that, while $n_1$ increases monotonically with $T_g$
and $\delta r_c$, the rate is somewhat smaller than that associated with the
preferential attachment rule [Fig.~\ref{fig:frequency}(b)], indicating
that the random attachment rule tends to make network collapses more frequent.
Figure~\ref{fig:rossler}(b) shows the algebraic distribution of the
collapse size: $p_{col}(\Delta n)\sim \Delta n^{\gamma}$ for
$\Delta n\in [1,50]$, with $\gamma\approx -0.58$. These results suggest
that the SOC characteristics of the network collapse phenomenon are robust,
regardless of the details of the network growth mechanism and of the
nodal dynamical processes.

For the randomly growing chaotic R\"{o}ssler network, we find that the
relationship between the synchronization error $\delta r_i$ and the
eigenvector element $e_{2,i}$ can still be described by \eqref{eq:delta_r_i}
[inset in Fig.~\ref{fig:rossler}(b)]. However, when analyzing the algebraic
size distribution using the eigenvectors, we note that the agreement between
the theoretical predication and the direct simulation results is not as
good as that for the preferential attachment growth rule. For example,
by truncating the eigenvector $\bm{e}_2$ of a random network of $n=800$
nodes, we obtain $p'_{col}(\Delta n) \sim \Delta n^{\gamma'}$ with
$\gamma'\approx -0.94$. The difference in the value of the algebraic
scaling exponent can be attributed to the limited size of the network
generated subject to the synchronization constraint as well as to the
relatively short transient period (small values of $T_g$). In fact,
in a computationally feasible implementation of the random growth model
with continuous-time dynamics, the largest network generated has the
size $n\approx 50$, rendering somewhat severe the finite size effect.
Nonetheless, in spite of the finite-size effect, the SOC features of the
network collapse phenomenon are robust.

\section{Conclusions} \label{sec:conclusion}

Growth or expansion is a fundamental feature of complex networks in
nature, society, and technological systems. Growth, however, is often
subject to constraints. Traditional models of complex networks contain
certain growth mechanism, such as one based on the
preferential attachment rule~\cite{BA:1999}, but impose no constraint.
Apparently, when growth is constrained, typically the network cannot
expand indefinitely, nor can its size be a monotonous function of time.
As a result, during the growth process there must be times when the
network size is reduced (collapse). But are there generic features
of the collapse events? For example, statistically what is the
distribution of the collapse size, and are there universal
characteristics in the distribution?

This paper addresses these intriguing questions using synchronization
as a concrete type of constraint. In particular, taking into account the
effects of desynchronization tolerance and synchronization speed, we
propose and investigate growing complex networks subject to the
constraint of synchronization stability. We find that, as new nodes
are continuously added into the network, it can self-organize itself
into a critical state where the addition of a single node can trigger
a large scale collapse. Statistical analysis of the characteristics
of the collapse events such as the degree distribution of the collapsed
nodes, the collapse frequency, and the collapse size distribution,
indicates that constraint induced network collapse can be viewed as
an evolutionary process towards self-organized criticality. The SOC
feature is especially pronounced as the collapse size follows an
algebraic scaling law. We develop an eigenvector analysis to understand the
origin of the network collapse phenomenon and the associated scaling
behaviors.

In a modern society, cities and infrastructures continue to expand.
In social media, various groups (social networks) keep growing. When
constraints are imposed, e.g., manifested as governmental policies or
online security rules, how would the underlying network respond? Can
constraints lead to large scale, catastrophic collapse of the entire
network? These are difficult but highly pertinent questions. Our findings
provide some hints about the dynamical features of the network collapse
phenomenon, but much further efforts are needed in this
direction of complex systems research.

\section*{Acknowledgement}

This work was supported by the National Natural Science Foundation of
China under Grant No.~11375109 and by the Fundamental Research Funds
for the Central Universities under Grant No.~GK201303002. YCL was
supported by ARO under Grant No.~W911NF-14-1-0504.

\bibliographystyle{apsrev4-1}
\bibliography{Network_Syn}

\begin{thebibliography}{81}%
\makeatletter
\providecommand \@ifxundefined [1]{%
 \@ifx{#1\undefined}
}%
\providecommand \@ifnum [1]{%
 \ifnum #1\expandafter \@firstoftwo
 \else \expandafter \@secondoftwo
 \fi
}%
\providecommand \@ifx [1]{%
 \ifx #1\expandafter \@firstoftwo
 \else \expandafter \@secondoftwo
 \fi
}%
\providecommand \natexlab [1]{#1}%
\providecommand \enquote  [1]{``#1''}%
\providecommand \bibnamefont  [1]{#1}%
\providecommand \bibfnamefont [1]{#1}%
\providecommand \citenamefont [1]{#1}%
\providecommand \href@noop [0]{\@secondoftwo}%
\providecommand \href [0]{\begingroup \@sanitize@url \@href}%
\providecommand \@href[1]{\@@startlink{#1}\@@href}%
\providecommand \@@href[1]{\endgroup#1\@@endlink}%
\providecommand \@sanitize@url [0]{\catcode `\\12\catcode `\$12\catcode
  `\&12\catcode `\#12\catcode `\^12\catcode `\_12\catcode `\%12\relax}%
\providecommand \@@startlink[1]{}%
\providecommand \@@endlink[0]{}%
\providecommand \url  [0]{\begingroup\@sanitize@url \@url }%
\providecommand \@url [1]{\endgroup\@href {#1}{\urlprefix }}%
\providecommand \urlprefix  [0]{URL }%
\providecommand \Eprint [0]{\href }%
\providecommand \doibase [0]{http://dx.doi.org/}%
\providecommand \selectlanguage [0]{\@gobble}%
\providecommand \bibinfo  [0]{\@secondoftwo}%
\providecommand \bibfield  [0]{\@secondoftwo}%
\providecommand \translation [1]{[#1]}%
\providecommand \BibitemOpen [0]{}%
\providecommand \bibitemStop [0]{}%
\providecommand \bibitemNoStop [0]{.\EOS\space}%
\providecommand \EOS [0]{\spacefactor3000\relax}%
\providecommand \BibitemShut  [1]{\csname bibitem#1\endcsname}%
\let\auto@bib@innerbib\@empty
\bibitem [{\citenamefont {Albert}\ and\ \citenamefont
  {Barab\'{a}si}(2002)}]{AB:2002}%
  \BibitemOpen
  \bibfield  {author} {\bibinfo {author} {\bibfnamefont {R.}~\bibnamefont
  {Albert}}\ and\ \bibinfo {author} {\bibfnamefont {A.-L.}\ \bibnamefont
  {Barab\'{a}si}},\ }\href@noop {} {\bibfield  {journal} {\bibinfo  {journal}
  {Rev. Mod. Phys.}\ }\textbf {\bibinfo {volume} {74}},\ \bibinfo {pages} {47}
  (\bibinfo {year} {2002})}\BibitemShut {NoStop}%
\bibitem [{\citenamefont {Newman}(2003)}]{Newman:2003}%
  \BibitemOpen
  \bibfield  {author} {\bibinfo {author} {\bibfnamefont {M.~E.~J.}\
  \bibnamefont {Newman}},\ }\href@noop {} {\bibfield  {journal} {\bibinfo
  {journal} {SIAM Rev.}\ }\textbf {\bibinfo {volume} {45}},\ \bibinfo {pages}
  {167} (\bibinfo {year} {2003})}\BibitemShut {NoStop}%
\bibitem [{\citenamefont {Barab\'{a}si}\ and\ \citenamefont
  {Albert}(1999)}]{BA:1999}%
  \BibitemOpen
  \bibfield  {author} {\bibinfo {author} {\bibfnamefont {A.-L.}\ \bibnamefont
  {Barab\'{a}si}}\ and\ \bibinfo {author} {\bibfnamefont {R.}~\bibnamefont
  {Albert}},\ }\href@noop {} {\bibfield  {journal} {\bibinfo  {journal}
  {Science}\ }\textbf {\bibinfo {volume} {286}},\ \bibinfo {pages} {509}
  (\bibinfo {year} {1999})}\BibitemShut {NoStop}%
\bibitem [{\citenamefont {Watts}(2002)}]{Watts:2002}%
  \BibitemOpen
  \bibfield  {author} {\bibinfo {author} {\bibfnamefont {D.~J.}\ \bibnamefont
  {Watts}},\ }\href@noop {} {\bibfield  {journal} {\bibinfo  {journal} {Proc.
  Natl. Acad. Sci. U.S.A.}\ }\textbf {\bibinfo {volume} {99}},\ \bibinfo
  {pages} {5766} (\bibinfo {year} {2002})}\BibitemShut {NoStop}%
\bibitem [{\citenamefont {Motter}\ and\ \citenamefont {Lai}(2002)}]{ML:2002}%
  \BibitemOpen
  \bibfield  {author} {\bibinfo {author} {\bibfnamefont {A.~E.}\ \bibnamefont
  {Motter}}\ and\ \bibinfo {author} {\bibfnamefont {Y.-C.}\ \bibnamefont
  {Lai}},\ }\href@noop {} {\bibfield  {journal} {\bibinfo  {journal} {Phys.
  Rev. E}\ }\textbf {\bibinfo {volume} {66}},\ \bibinfo {pages} {065102(R)}
  (\bibinfo {year} {2002})}\BibitemShut {NoStop}%
\bibitem [{\citenamefont {Holme}\ and\ \citenamefont {Kim}(2002)}]{HK:2002}%
  \BibitemOpen
  \bibfield  {author} {\bibinfo {author} {\bibfnamefont {P.}~\bibnamefont
  {Holme}}\ and\ \bibinfo {author} {\bibfnamefont {B.~J.}\ \bibnamefont
  {Kim}},\ }\href@noop {} {\bibfield  {journal} {\bibinfo  {journal} {Phys.
  Rev. E}\ }\textbf {\bibinfo {volume} {65}},\ \bibinfo {pages} {066109}
  (\bibinfo {year} {2002})}\BibitemShut {NoStop}%
\bibitem [{\citenamefont {Moreno}\ \emph {et~al.}(2002)\citenamefont {Moreno},
  \citenamefont {G\'{o}mez},\ and\ \citenamefont {Pacheco}}]{MGP:2002}%
  \BibitemOpen
  \bibfield  {author} {\bibinfo {author} {\bibfnamefont {Y.}~\bibnamefont
  {Moreno}}, \bibinfo {author} {\bibfnamefont {J.~B.}\ \bibnamefont
  {G\'{o}mez}}, \ and\ \bibinfo {author} {\bibfnamefont {A.~F.}\ \bibnamefont
  {Pacheco}},\ }\href@noop {} {\bibfield  {journal} {\bibinfo  {journal}
  {Europhys. Lett.}\ }\textbf {\bibinfo {volume} {58}},\ \bibinfo {pages} {630}
  (\bibinfo {year} {2002})}\BibitemShut {NoStop}%
\bibitem [{\citenamefont {Moreno}\ \emph {et~al.}(2003)\citenamefont {Moreno},
  \citenamefont {Pastor-Satorras}, \citenamefont {V\'{a}zquez},\ and\
  \citenamefont {Vespignani}}]{MPSVV:2003}%
  \BibitemOpen
  \bibfield  {author} {\bibinfo {author} {\bibfnamefont {Y.}~\bibnamefont
  {Moreno}}, \bibinfo {author} {\bibfnamefont {R.}~\bibnamefont
  {Pastor-Satorras}}, \bibinfo {author} {\bibfnamefont {A.}~\bibnamefont
  {V\'{a}zquez}}, \ and\ \bibinfo {author} {\bibfnamefont {A.}~\bibnamefont
  {Vespignani}},\ }\href@noop {} {\bibfield  {journal} {\bibinfo  {journal}
  {Europhys. Lett.}\ }\textbf {\bibinfo {volume} {62}},\ \bibinfo {pages} {292}
  (\bibinfo {year} {2003})}\BibitemShut {NoStop}%
\bibitem [{\citenamefont {Holme}(2002)}]{Holme:2002}%
  \BibitemOpen
  \bibfield  {author} {\bibinfo {author} {\bibfnamefont {P.}~\bibnamefont
  {Holme}},\ }\href@noop {} {\bibfield  {journal} {\bibinfo  {journal} {Phys.
  Rev. E}\ }\textbf {\bibinfo {volume} {66}},\ \bibinfo {pages} {036119}
  (\bibinfo {year} {2002})}\BibitemShut {NoStop}%
\bibitem [{\citenamefont {Goh}\ \emph {et~al.}(2003)\citenamefont {Goh},
  \citenamefont {Lee}, \citenamefont {Kahng},\ and\ \citenamefont
  {Kim}}]{GLKK:2003}%
  \BibitemOpen
  \bibfield  {author} {\bibinfo {author} {\bibfnamefont {K.-I.}\ \bibnamefont
  {Goh}}, \bibinfo {author} {\bibfnamefont {D.-S.}\ \bibnamefont {Lee}},
  \bibinfo {author} {\bibfnamefont {B.}~\bibnamefont {Kahng}}, \ and\ \bibinfo
  {author} {\bibfnamefont {D.}~\bibnamefont {Kim}},\ }\href@noop {} {\bibfield
  {journal} {\bibinfo  {journal} {Phys. Rev. Lett.}\ }\textbf {\bibinfo
  {volume} {93}},\ \bibinfo {pages} {148701} (\bibinfo {year}
  {2003})}\BibitemShut {NoStop}%
\bibitem [{\citenamefont {Crucitti}\ \emph {et~al.}(2004)\citenamefont
  {Crucitti}, \citenamefont {Latora},\ and\ \citenamefont
  {Marchior}}]{CLM:2004}%
  \BibitemOpen
  \bibfield  {author} {\bibinfo {author} {\bibfnamefont {P.}~\bibnamefont
  {Crucitti}}, \bibinfo {author} {\bibfnamefont {V.}~\bibnamefont {Latora}}, \
  and\ \bibinfo {author} {\bibfnamefont {M.}~\bibnamefont {Marchior}},\
  }\href@noop {} {\bibfield  {journal} {\bibinfo  {journal} {Phys. Rev. E}\
  }\textbf {\bibinfo {volume} {69}},\ \bibinfo {pages} {045104(R)} (\bibinfo
  {year} {2004})}\BibitemShut {NoStop}%
\bibitem [{\citenamefont {Huang}\ \emph
  {et~al.}(2006{\natexlab{a}})\citenamefont {Huang}, \citenamefont {Yang},\
  and\ \citenamefont {Yang}}]{HYY:2006}%
  \BibitemOpen
  \bibfield  {author} {\bibinfo {author} {\bibfnamefont {L.}~\bibnamefont
  {Huang}}, \bibinfo {author} {\bibfnamefont {L.}~\bibnamefont {Yang}}, \ and\
  \bibinfo {author} {\bibfnamefont {K.-Q.}\ \bibnamefont {Yang}},\ }\href@noop
  {} {\bibfield  {journal} {\bibinfo  {journal} {Phys. Rev. E}\ }\textbf
  {\bibinfo {volume} {73}},\ \bibinfo {pages} {036102} (\bibinfo {year}
  {2006}{\natexlab{a}})}\BibitemShut {NoStop}%
\bibitem [{\citenamefont {Galstyan}\ and\ \citenamefont
  {Cohen}(2007)}]{GC:2007}%
  \BibitemOpen
  \bibfield  {author} {\bibinfo {author} {\bibfnamefont {A.}~\bibnamefont
  {Galstyan}}\ and\ \bibinfo {author} {\bibfnamefont {P.}~\bibnamefont
  {Cohen}},\ }\href@noop {} {\bibfield  {journal} {\bibinfo  {journal} {Phys.
  Rev. E}\ }\textbf {\bibinfo {volume} {75}},\ \bibinfo {pages} {036109}
  (\bibinfo {year} {2007})}\BibitemShut {NoStop}%
\bibitem [{\citenamefont {Huang}\ \emph
  {et~al.}(2008{\natexlab{a}})\citenamefont {Huang}, \citenamefont {Lai},\ and\
  \citenamefont {Chen}}]{HLC:2008}%
  \BibitemOpen
  \bibfield  {author} {\bibinfo {author} {\bibfnamefont {L.}~\bibnamefont
  {Huang}}, \bibinfo {author} {\bibfnamefont {Y.-C.}\ \bibnamefont {Lai}}, \
  and\ \bibinfo {author} {\bibfnamefont {G.}~\bibnamefont {Chen}},\ }\href@noop
  {} {\bibfield  {journal} {\bibinfo  {journal} {Phys. Rev. E}\ }\textbf
  {\bibinfo {volume} {78}},\ \bibinfo {pages} {036116} (\bibinfo {year}
  {2008}{\natexlab{a}})}\BibitemShut {NoStop}%
\bibitem [{\citenamefont {Simonsen}\ \emph {et~al.}(2008)\citenamefont
  {Simonsen}, \citenamefont {Buzna}, \citenamefont {Peters}, \citenamefont
  {Bornholdt},\ and\ \citenamefont {Helbing}}]{SBPBH:2008}%
  \BibitemOpen
  \bibfield  {author} {\bibinfo {author} {\bibfnamefont {I.}~\bibnamefont
  {Simonsen}}, \bibinfo {author} {\bibfnamefont {L.}~\bibnamefont {Buzna}},
  \bibinfo {author} {\bibfnamefont {K.}~\bibnamefont {Peters}}, \bibinfo
  {author} {\bibfnamefont {S.}~\bibnamefont {Bornholdt}}, \ and\ \bibinfo
  {author} {\bibfnamefont {D.}~\bibnamefont {Helbing}},\ }\href@noop {}
  {\bibfield  {journal} {\bibinfo  {journal} {Phys. Rev. Lett.}\ }\textbf
  {\bibinfo {volume} {100}},\ \bibinfo {pages} {218701} (\bibinfo {year}
  {2008})}\BibitemShut {NoStop}%
\bibitem [{\citenamefont {Gleeson}(2008)}]{Gleeson:2008}%
  \BibitemOpen
  \bibfield  {author} {\bibinfo {author} {\bibfnamefont {J.~P.}\ \bibnamefont
  {Gleeson}},\ }\href@noop {} {\bibfield  {journal} {\bibinfo  {journal} {Phys.
  Rev. E}\ }\textbf {\bibinfo {volume} {77}},\ \bibinfo {pages} {046117}
  (\bibinfo {year} {2008})}\BibitemShut {NoStop}%
\bibitem [{\citenamefont {Yang}\ \emph {et~al.}(2009)\citenamefont {Yang},
  \citenamefont {Wang}, \citenamefont {Lai},\ and\ \citenamefont
  {Chen}}]{YWLC:2009}%
  \BibitemOpen
  \bibfield  {author} {\bibinfo {author} {\bibfnamefont {R.}~\bibnamefont
  {Yang}}, \bibinfo {author} {\bibfnamefont {W.-X.}\ \bibnamefont {Wang}},
  \bibinfo {author} {\bibfnamefont {Y.-C.}\ \bibnamefont {Lai}}, \ and\
  \bibinfo {author} {\bibfnamefont {G.}~\bibnamefont {Chen}},\ }\href@noop {}
  {\bibfield  {journal} {\bibinfo  {journal} {Phys. Rev. E}\ }\textbf {\bibinfo
  {volume} {79}},\ \bibinfo {pages} {026112} (\bibinfo {year}
  {2009})}\BibitemShut {NoStop}%
\bibitem [{\citenamefont {Whitney}(2010)}]{Whitney:2010}%
  \BibitemOpen
  \bibfield  {author} {\bibinfo {author} {\bibfnamefont {D.~E.}\ \bibnamefont
  {Whitney}},\ }\href@noop {} {\bibfield  {journal} {\bibinfo  {journal} {Phys.
  Rev. E}\ }\textbf {\bibinfo {volume} {82}},\ \bibinfo {pages} {066110}
  (\bibinfo {year} {2010})}\BibitemShut {NoStop}%
\bibitem [{\citenamefont {Wang}\ \emph {et~al.}(2010)\citenamefont {Wang},
  \citenamefont {Yang},\ and\ \citenamefont {Lai}}]{WYL:2010}%
  \BibitemOpen
  \bibfield  {author} {\bibinfo {author} {\bibfnamefont {W.-X.}\ \bibnamefont
  {Wang}}, \bibinfo {author} {\bibfnamefont {R.}~\bibnamefont {Yang}}, \ and\
  \bibinfo {author} {\bibfnamefont {Y.-C.}\ \bibnamefont {Lai}},\ }\href@noop
  {} {\bibfield  {journal} {\bibinfo  {journal} {Phys. Rev. E}\ }\textbf
  {\bibinfo {volume} {81}},\ \bibinfo {pages} {035102(R)} (\bibinfo {year}
  {2010})}\BibitemShut {NoStop}%
\bibitem [{\citenamefont {Huang}\ and\ \citenamefont {Lai}(2011)}]{HL:2011}%
  \BibitemOpen
  \bibfield  {author} {\bibinfo {author} {\bibfnamefont {L.}~\bibnamefont
  {Huang}}\ and\ \bibinfo {author} {\bibfnamefont {Y.-C.}\ \bibnamefont
  {Lai}},\ }\href@noop {} {\bibfield  {journal} {\bibinfo  {journal} {Chaos}\
  }\textbf {\bibinfo {volume} {21}},\ \bibinfo {pages} {025107} (\bibinfo
  {year} {2011})}\BibitemShut {NoStop}%
\bibitem [{\citenamefont {Wang}\ \emph {et~al.}(2011)\citenamefont {Wang},
  \citenamefont {Lai},\ and\ \citenamefont {Armbruster}}]{WLA:2011}%
  \BibitemOpen
  \bibfield  {author} {\bibinfo {author} {\bibfnamefont {W.-X.}\ \bibnamefont
  {Wang}}, \bibinfo {author} {\bibfnamefont {Y.-C.}\ \bibnamefont {Lai}}, \
  and\ \bibinfo {author} {\bibfnamefont {D.}~\bibnamefont {Armbruster}},\
  }\href@noop {} {\bibfield  {journal} {\bibinfo  {journal} {Chaos}\ }\textbf
  {\bibinfo {volume} {21}},\ \bibinfo {pages} {033112} (\bibinfo {year}
  {2011})}\BibitemShut {NoStop}%
\bibitem [{\citenamefont {Liu}\ \emph {et~al.}(2012)\citenamefont {Liu},
  \citenamefont {Wang}, \citenamefont {Lai},\ and\ \citenamefont
  {Wang}}]{LWLW:2012}%
  \BibitemOpen
  \bibfield  {author} {\bibinfo {author} {\bibfnamefont {R.-R.}\ \bibnamefont
  {Liu}}, \bibinfo {author} {\bibfnamefont {W.-X.}\ \bibnamefont {Wang}},
  \bibinfo {author} {\bibfnamefont {Y.-C.}\ \bibnamefont {Lai}}, \ and\
  \bibinfo {author} {\bibfnamefont {B.-H.}\ \bibnamefont {Wang}},\ }\href@noop
  {} {\bibfield  {journal} {\bibinfo  {journal} {Phys. Rev. E}\ }\textbf
  {\bibinfo {volume} {85}},\ \bibinfo {pages} {026110} (\bibinfo {year}
  {2012})}\BibitemShut {NoStop}%
\bibitem [{\citenamefont {Helbing}(2013)}]{Helbing:2013}%
  \BibitemOpen
  \bibfield  {author} {\bibinfo {author} {\bibfnamefont {D.}~\bibnamefont
  {Helbing}},\ }\href@noop {} {\bibfield  {journal} {\bibinfo  {journal}
  {Nature (London)}\ }\textbf {\bibinfo {volume} {497}},\ \bibinfo {pages} {51}
  (\bibinfo {year} {2013})}\BibitemShut {NoStop}%
\bibitem [{\citenamefont {Gajduk}\ \emph {et~al.}(2014)\citenamefont {Gajduk},
  \citenamefont {Todorovsky},\ and\ \citenamefont {Kocarev}}]{GTK:2014}%
  \BibitemOpen
  \bibfield  {author} {\bibinfo {author} {\bibfnamefont {A.}~\bibnamefont
  {Gajduk}}, \bibinfo {author} {\bibfnamefont {M.}~\bibnamefont {Todorovsky}},
  \ and\ \bibinfo {author} {\bibfnamefont {L.}~\bibnamefont {Kocarev}},\
  }\href@noop {} {\bibfield  {journal} {\bibinfo  {journal} {Eur. Phys. J. Spe.
  Top.}\ }\textbf {\bibinfo {volume} {223}},\ \bibinfo {pages} {2387} (\bibinfo
  {year} {2014})}\BibitemShut {NoStop}%
\bibitem [{\citenamefont {May}(1972)}]{May:1972}%
  \BibitemOpen
  \bibfield  {author} {\bibinfo {author} {\bibfnamefont {R.~M.}\ \bibnamefont
  {May}},\ }\href@noop {} {\bibfield  {journal} {\bibinfo  {journal} {Nature
  (London)}\ }\textbf {\bibinfo {volume} {238}},\ \bibinfo {pages} {413}
  (\bibinfo {year} {1972})}\BibitemShut {NoStop}%
\bibitem [{\citenamefont {Proulx}\ \emph {et~al.}(2005)\citenamefont {Proulx},
  \citenamefont {Promislow},\ and\ \citenamefont {Phillips}}]{PPP:2005}%
  \BibitemOpen
  \bibfield  {author} {\bibinfo {author} {\bibfnamefont {S.~R.}\ \bibnamefont
  {Proulx}}, \bibinfo {author} {\bibfnamefont {D.~E.~L.}\ \bibnamefont
  {Promislow}}, \ and\ \bibinfo {author} {\bibfnamefont {P.~C.}\ \bibnamefont
  {Phillips}},\ }\href@noop {} {\bibfield  {journal} {\bibinfo  {journal}
  {Trends Ecol. Evol.}\ }\textbf {\bibinfo {volume} {20}},\ \bibinfo {pages}
  {345} (\bibinfo {year} {2005})}\BibitemShut {NoStop}%
\bibitem [{\citenamefont {Commission}(2011)}]{Financial_Crisis}%
  \BibitemOpen
  \bibfield  {author} {\bibinfo {author} {\bibfnamefont {T.~F. C.~I.}\
  \bibnamefont {Commission}},\ }\href@noop {} {\emph {\bibinfo {title} {The
  Financial Crisis Inquiry Report: Final Report of the National Commission on
  the Causes of the Current Financial and Economic Crisis in the United
  States}}}\ (\bibinfo  {publisher} {Public Affairs},\ \bibinfo {year}
  {2011})\BibitemShut {NoStop}%
\bibitem [{\citenamefont {Perotti}\ \emph {et~al.}(2009)\citenamefont
  {Perotti}, \citenamefont {Billoni}, \citenamefont {Tamarit}, \citenamefont
  {Chialvo},\ and\ \citenamefont {Cannas}}]{PBTCC:2009}%
  \BibitemOpen
  \bibfield  {author} {\bibinfo {author} {\bibfnamefont {J.~I.}\ \bibnamefont
  {Perotti}}, \bibinfo {author} {\bibfnamefont {O.~V.}\ \bibnamefont
  {Billoni}}, \bibinfo {author} {\bibfnamefont {F.~A.}\ \bibnamefont
  {Tamarit}}, \bibinfo {author} {\bibfnamefont {D.~R.}\ \bibnamefont
  {Chialvo}}, \ and\ \bibinfo {author} {\bibfnamefont {S.~A.}\ \bibnamefont
  {Cannas}},\ }\href {\doibase 10.1103/PhysRevLett.103.108701} {\bibfield
  {journal} {\bibinfo  {journal} {Phys. Rev. Lett.}\ }\textbf {\bibinfo
  {volume} {103}},\ \bibinfo {pages} {108701} (\bibinfo {year}
  {2009})}\BibitemShut {NoStop}%
\bibitem [{\citenamefont {Fu}\ and\ \citenamefont {Wang}(2011)}]{FW:2011}%
  \BibitemOpen
  \bibfield  {author} {\bibinfo {author} {\bibfnamefont {C.}~\bibnamefont
  {Fu}}\ and\ \bibinfo {author} {\bibfnamefont {X.}~\bibnamefont {Wang}},\
  }\href {\doibase 10.1103/PhysRevE.83.066101} {\bibfield  {journal} {\bibinfo
  {journal} {Phys. Rev. E}\ }\textbf {\bibinfo {volume} {83}},\ \bibinfo
  {pages} {066101} (\bibinfo {year} {2011})}\BibitemShut {NoStop}%
\bibitem [{\citenamefont {Kuramoto}(1984)}]{Kuramoto:book}%
  \BibitemOpen
  \bibfield  {author} {\bibinfo {author} {\bibfnamefont {Y.}~\bibnamefont
  {Kuramoto}},\ }\href@noop {} {\emph {\bibinfo {title} {Chemical Oscillations,
  Waves and Turbulence}}},\ \bibinfo {edition} {1st}\ ed.\ (\bibinfo
  {publisher} {Springer-Verlag},\ \bibinfo {address} {Berlin},\ \bibinfo {year}
  {1984})\BibitemShut {NoStop}%
\bibitem [{\citenamefont {Strogatz}(2003)}]{Strogatz:book}%
  \BibitemOpen
  \bibfield  {author} {\bibinfo {author} {\bibfnamefont {S.}~\bibnamefont
  {Strogatz}},\ }\href@noop {} {\emph {\bibinfo {title} {Sync: The Emerging
  Science of Spontaneous Order}}},\ \bibinfo {edition} {1st}\ ed.\ (\bibinfo
  {publisher} {Hyperion},\ \bibinfo {address} {New York},\ \bibinfo {year}
  {2003})\BibitemShut {NoStop}%
\bibitem [{\citenamefont {Pikovsky}\ \emph {et~al.}(2001)\citenamefont
  {Pikovsky}, \citenamefont {Rosenblum},\ and\ \citenamefont
  {Kurths}}]{PRK:book}%
  \BibitemOpen
  \bibfield  {author} {\bibinfo {author} {\bibfnamefont {A.~S.}\ \bibnamefont
  {Pikovsky}}, \bibinfo {author} {\bibfnamefont {M.~G.}\ \bibnamefont
  {Rosenblum}}, \ and\ \bibinfo {author} {\bibfnamefont {J.}~\bibnamefont
  {Kurths}},\ }\href@noop {} {\emph {\bibinfo {title} {Synchronization: A
  universal concept in nonlinear sciences}}},\ \bibinfo {edition} {1st}\ ed.\
  (\bibinfo  {publisher} {Cambridge University Press},\ \bibinfo {address}
  {Cambridge},\ \bibinfo {year} {2001})\BibitemShut {NoStop}%
\bibitem [{\citenamefont {Fujisaka}\ and\ \citenamefont
  {Yamada}(1983)}]{FY:1983}%
  \BibitemOpen
  \bibfield  {author} {\bibinfo {author} {\bibfnamefont {H.}~\bibnamefont
  {Fujisaka}}\ and\ \bibinfo {author} {\bibfnamefont {T.}~\bibnamefont
  {Yamada}},\ }\href@noop {} {\bibfield  {journal} {\bibinfo  {journal} {Prog.
  Theor. Phys.}\ }\textbf {\bibinfo {volume} {69}},\ \bibinfo {pages} {32}
  (\bibinfo {year} {1983})}\BibitemShut {NoStop}%
\bibitem [{\citenamefont {Pecora}\ and\ \citenamefont
  {Carroll}(1990)}]{PC:1990}%
  \BibitemOpen
  \bibfield  {author} {\bibinfo {author} {\bibfnamefont {L.~M.}\ \bibnamefont
  {Pecora}}\ and\ \bibinfo {author} {\bibfnamefont {T.~L.}\ \bibnamefont
  {Carroll}},\ }\href@noop {} {\bibfield  {journal} {\bibinfo  {journal} {Phys.
  Rev. Lett.}\ }\textbf {\bibinfo {volume} {64}},\ \bibinfo {pages} {821}
  (\bibinfo {year} {1990})}\BibitemShut {NoStop}%
\bibitem [{\citenamefont {Arenas}\ \emph {et~al.}(2008)\citenamefont {Arenas},
  \citenamefont {Diaz-Guilera}, \citenamefont {Kurths}, \citenamefont
  {Morenob},\ and\ \citenamefont {Zhou}}]{ADGKMZ:2008}%
  \BibitemOpen
  \bibfield  {author} {\bibinfo {author} {\bibfnamefont {A.}~\bibnamefont
  {Arenas}}, \bibinfo {author} {\bibfnamefont {A.}~\bibnamefont
  {Diaz-Guilera}}, \bibinfo {author} {\bibfnamefont {J.}~\bibnamefont
  {Kurths}}, \bibinfo {author} {\bibfnamefont {Y.}~\bibnamefont {Morenob}}, \
  and\ \bibinfo {author} {\bibfnamefont {C.-S.}\ \bibnamefont {Zhou}},\
  }\href@noop {} {\bibfield  {journal} {\bibinfo  {journal} {Phys. Rep.}\
  }\textbf {\bibinfo {volume} {469}},\ \bibinfo {pages} {93} (\bibinfo {year}
  {2008})}\BibitemShut {NoStop}%
\bibitem [{\citenamefont {Watts}\ and\ \citenamefont
  {Strogatz}(1998)}]{WS:1998}%
  \BibitemOpen
  \bibfield  {author} {\bibinfo {author} {\bibfnamefont {D.~J.}\ \bibnamefont
  {Watts}}\ and\ \bibinfo {author} {\bibfnamefont {S.~H.}\ \bibnamefont
  {Strogatz}},\ }\href@noop {} {\bibfield  {journal} {\bibinfo  {journal}
  {Nature (London)}\ }\textbf {\bibinfo {volume} {393}},\ \bibinfo {pages}
  {440} (\bibinfo {year} {1998})}\BibitemShut {NoStop}%
\bibitem [{\citenamefont {Barahona}\ and\ \citenamefont
  {Pecora}(2002)}]{BP:2002}%
  \BibitemOpen
  \bibfield  {author} {\bibinfo {author} {\bibfnamefont {M.}~\bibnamefont
  {Barahona}}\ and\ \bibinfo {author} {\bibfnamefont {L.~M.}\ \bibnamefont
  {Pecora}},\ }\href@noop {} {\bibfield  {journal} {\bibinfo  {journal} {Phys.
  Rev. Lett.}\ }\textbf {\bibinfo {volume} {89}},\ \bibinfo {pages} {054101}
  (\bibinfo {year} {2002})}\BibitemShut {NoStop}%
\bibitem [{\citenamefont {Nishikawa}\ \emph {et~al.}(2003)\citenamefont
  {Nishikawa}, \citenamefont {Motter}, \citenamefont {Lai},\ and\ \citenamefont
  {Hoppensteadt}}]{NMLH:2003}%
  \BibitemOpen
  \bibfield  {author} {\bibinfo {author} {\bibfnamefont {T.}~\bibnamefont
  {Nishikawa}}, \bibinfo {author} {\bibfnamefont {A.~E.}\ \bibnamefont
  {Motter}}, \bibinfo {author} {\bibfnamefont {Y.-C.}\ \bibnamefont {Lai}}, \
  and\ \bibinfo {author} {\bibfnamefont {F.}~\bibnamefont {Hoppensteadt}},\
  }\href@noop {} {\bibfield  {journal} {\bibinfo  {journal} {Phys. Rev. Lett.}\
  }\textbf {\bibinfo {volume} {91}},\ \bibinfo {pages} {014101} (\bibinfo
  {year} {2003})}\BibitemShut {NoStop}%
\bibitem [{\citenamefont {Donetti}\ \emph {et~al.}(2005)\citenamefont
  {Donetti}, \citenamefont {Hurtado},\ and\ \citenamefont
  {Mu\~noz}}]{DHM:2005}%
  \BibitemOpen
  \bibfield  {author} {\bibinfo {author} {\bibfnamefont {L.}~\bibnamefont
  {Donetti}}, \bibinfo {author} {\bibfnamefont {P.~I.}\ \bibnamefont
  {Hurtado}}, \ and\ \bibinfo {author} {\bibfnamefont {M.~A.}\ \bibnamefont
  {Mu\~noz}},\ }\href {\doibase 10.1103/PhysRevLett.95.188701} {\bibfield
  {journal} {\bibinfo  {journal} {Phys. Rev. Lett.}\ }\textbf {\bibinfo
  {volume} {95}},\ \bibinfo {pages} {188701} (\bibinfo {year}
  {2005})}\BibitemShut {NoStop}%
\bibitem [{\citenamefont {Belykh}\ \emph {et~al.}(2005)\citenamefont {Belykh},
  \citenamefont {Hasler}, \citenamefont {Lauret},\ and\ \citenamefont
  {Nijmeijer}}]{BHLN:2005}%
  \BibitemOpen
  \bibfield  {author} {\bibinfo {author} {\bibfnamefont {I.}~\bibnamefont
  {Belykh}}, \bibinfo {author} {\bibfnamefont {M.}~\bibnamefont {Hasler}},
  \bibinfo {author} {\bibfnamefont {M.}~\bibnamefont {Lauret}}, \ and\ \bibinfo
  {author} {\bibfnamefont {H.}~\bibnamefont {Nijmeijer}},\ }\href@noop {}
  {\bibfield  {journal} {\bibinfo  {journal} {Int. J. Bif. Chaos}\ }\textbf
  {\bibinfo {volume} {15}},\ \bibinfo {pages} {3423} (\bibinfo {year}
  {2005})}\BibitemShut {NoStop}%
\bibitem [{\citenamefont {Oh}\ \emph {et~al.}(2005)\citenamefont {Oh},
  \citenamefont {Rho}, \citenamefont {Hong},\ and\ \citenamefont
  {Kahng}}]{ORHK:2005}%
  \BibitemOpen
  \bibfield  {author} {\bibinfo {author} {\bibfnamefont {E.}~\bibnamefont
  {Oh}}, \bibinfo {author} {\bibfnamefont {K.}~\bibnamefont {Rho}}, \bibinfo
  {author} {\bibfnamefont {H.}~\bibnamefont {Hong}}, \ and\ \bibinfo {author}
  {\bibfnamefont {B.}~\bibnamefont {Kahng}},\ }\href@noop {} {\bibfield
  {journal} {\bibinfo  {journal} {Phys. Rev. E}\ }\textbf {\bibinfo {volume}
  {72}},\ \bibinfo {pages} {047101} (\bibinfo {year} {2005})}\BibitemShut
  {NoStop}%
\bibitem [{\citenamefont {Restrepo}\ \emph {et~al.}(2005)\citenamefont
  {Restrepo}, \citenamefont {Ott},\ and\ \citenamefont {Hunt}}]{ROH:2005}%
  \BibitemOpen
  \bibfield  {author} {\bibinfo {author} {\bibfnamefont {J.~G.}\ \bibnamefont
  {Restrepo}}, \bibinfo {author} {\bibfnamefont {E.}~\bibnamefont {Ott}}, \
  and\ \bibinfo {author} {\bibfnamefont {B.~R.}\ \bibnamefont {Hunt}},\
  }\href@noop {} {\bibfield  {journal} {\bibinfo  {journal} {Phys. Rev. E}\
  }\textbf {\bibinfo {volume} {71}},\ \bibinfo {pages} {036151} (\bibinfo
  {year} {2005})}\BibitemShut {NoStop}%
\bibitem [{\citenamefont {Huang}\ \emph
  {et~al.}(2006{\natexlab{b}})\citenamefont {Huang}, \citenamefont {Park},
  \citenamefont {Lai}, \citenamefont {Yang},\ and\ \citenamefont
  {Yang}}]{HPLYY:2006}%
  \BibitemOpen
  \bibfield  {author} {\bibinfo {author} {\bibfnamefont {L.}~\bibnamefont
  {Huang}}, \bibinfo {author} {\bibfnamefont {K.}~\bibnamefont {Park}},
  \bibinfo {author} {\bibfnamefont {Y.-C.}\ \bibnamefont {Lai}}, \bibinfo
  {author} {\bibfnamefont {L.}~\bibnamefont {Yang}}, \ and\ \bibinfo {author}
  {\bibfnamefont {K.}~\bibnamefont {Yang}},\ }\href {\doibase
  10.1103/PhysRevLett.97.164101} {\bibfield  {journal} {\bibinfo  {journal}
  {Phys. Rev. Lett.}\ }\textbf {\bibinfo {volume} {97}},\ \bibinfo {pages}
  {164101} (\bibinfo {year} {2006}{\natexlab{b}})}\BibitemShut {NoStop}%
\bibitem [{\citenamefont {Belykh}\ \emph {et~al.}(2006)\citenamefont {Belykh},
  \citenamefont {Belykh},\ and\ \citenamefont {Hasler}}]{BBH:2006}%
  \BibitemOpen
  \bibfield  {author} {\bibinfo {author} {\bibfnamefont {V.}~\bibnamefont
  {Belykh}}, \bibinfo {author} {\bibfnamefont {I.}~\bibnamefont {Belykh}}, \
  and\ \bibinfo {author} {\bibfnamefont {M.}~\bibnamefont {Hasler}},\
  }\href@noop {} {\bibfield  {journal} {\bibinfo  {journal} {Physica D}\
  }\textbf {\bibinfo {volume} {224}},\ \bibinfo {pages} {42} (\bibinfo {year}
  {2006})}\BibitemShut {NoStop}%
\bibitem [{\citenamefont {Wang}\ \emph
  {et~al.}(2007{\natexlab{a}})\citenamefont {Wang}, \citenamefont {Huang},
  \citenamefont {Lai},\ and\ \citenamefont {Lai}}]{WHLL:2007}%
  \BibitemOpen
  \bibfield  {author} {\bibinfo {author} {\bibfnamefont {X.~G.}\ \bibnamefont
  {Wang}}, \bibinfo {author} {\bibfnamefont {L.}~\bibnamefont {Huang}},
  \bibinfo {author} {\bibfnamefont {Y.-C.}\ \bibnamefont {Lai}}, \ and\
  \bibinfo {author} {\bibfnamefont {C.-H.}\ \bibnamefont {Lai}},\ }\href@noop
  {} {\bibfield  {journal} {\bibinfo  {journal} {Phys. Rev. E}\ }\textbf
  {\bibinfo {volume} {76}},\ \bibinfo {pages} {056113} (\bibinfo {year}
  {2007}{\natexlab{a}})}\BibitemShut {NoStop}%
\bibitem [{\citenamefont {G\'omez-Garde\~nes}\ \emph
  {et~al.}(2007)\citenamefont {G\'omez-Garde\~nes}, \citenamefont {Moreno},\
  and\ \citenamefont {Arenas}}]{GMA:2007}%
  \BibitemOpen
  \bibfield  {author} {\bibinfo {author} {\bibfnamefont {J.}~\bibnamefont
  {G\'omez-Garde\~nes}}, \bibinfo {author} {\bibfnamefont {Y.}~\bibnamefont
  {Moreno}}, \ and\ \bibinfo {author} {\bibfnamefont {A.}~\bibnamefont
  {Arenas}},\ }\href {\doibase 10.1103/PhysRevLett.98.034101} {\bibfield
  {journal} {\bibinfo  {journal} {Phys. Rev. Lett.}\ }\textbf {\bibinfo
  {volume} {98}},\ \bibinfo {pages} {034101} (\bibinfo {year}
  {2007})}\BibitemShut {NoStop}%
\bibitem [{\citenamefont {Guan}\ \emph {et~al.}(2008)\citenamefont {Guan},
  \citenamefont {Wang}, \citenamefont {Lai},\ and\ \citenamefont
  {Lai}}]{GWLL:2008}%
  \BibitemOpen
  \bibfield  {author} {\bibinfo {author} {\bibfnamefont {S.-G.}\ \bibnamefont
  {Guan}}, \bibinfo {author} {\bibfnamefont {X.-G.}\ \bibnamefont {Wang}},
  \bibinfo {author} {\bibfnamefont {Y.-C.}\ \bibnamefont {Lai}}, \ and\
  \bibinfo {author} {\bibfnamefont {C.~H.}\ \bibnamefont {Lai}},\ }\href@noop
  {} {\bibfield  {journal} {\bibinfo  {journal} {Phys. Rev. E}\ }\textbf
  {\bibinfo {volume} {77}},\ \bibinfo {pages} {046211} (\bibinfo {year}
  {2008})}\BibitemShut {NoStop}%
\bibitem [{\citenamefont {Huang}\ \emph
  {et~al.}(2008{\natexlab{b}})\citenamefont {Huang}, \citenamefont {Lai},\ and\
  \citenamefont {Gatenby}}]{HLG:2008a}%
  \BibitemOpen
  \bibfield  {author} {\bibinfo {author} {\bibfnamefont {L.}~\bibnamefont
  {Huang}}, \bibinfo {author} {\bibfnamefont {Y.-C.}\ \bibnamefont {Lai}}, \
  and\ \bibinfo {author} {\bibfnamefont {R.~A.}\ \bibnamefont {Gatenby}},\
  }\href@noop {} {\bibfield  {journal} {\bibinfo  {journal} {Chaos}\ }\textbf
  {\bibinfo {volume} {18}},\ \bibinfo {pages} {013101} (\bibinfo {year}
  {2008}{\natexlab{b}})}\BibitemShut {NoStop}%
\bibitem [{\citenamefont {Huang}\ \emph
  {et~al.}(2008{\natexlab{c}})\citenamefont {Huang}, \citenamefont {Lai},\ and\
  \citenamefont {Gatenby}}]{HLG:2008b}%
  \BibitemOpen
  \bibfield  {author} {\bibinfo {author} {\bibfnamefont {L.}~\bibnamefont
  {Huang}}, \bibinfo {author} {\bibfnamefont {Y.-C.}\ \bibnamefont {Lai}}, \
  and\ \bibinfo {author} {\bibfnamefont {R.~A.}\ \bibnamefont {Gatenby}},\
  }\href@noop {} {\bibfield  {journal} {\bibinfo  {journal} {Phys. Rev. E}\
  }\textbf {\bibinfo {volume} {77}},\ \bibinfo {pages} {016103} (\bibinfo
  {year} {2008}{\natexlab{c}})}\BibitemShut {NoStop}%
\bibitem [{\citenamefont {Wang}\ \emph {et~al.}(2008)\citenamefont {Wang},
  \citenamefont {Huang}, \citenamefont {Guan}, \citenamefont {Lai},\ and\
  \citenamefont {Lai}}]{WHGLL:2008}%
  \BibitemOpen
  \bibfield  {author} {\bibinfo {author} {\bibfnamefont {X.-G.}\ \bibnamefont
  {Wang}}, \bibinfo {author} {\bibfnamefont {L.}~\bibnamefont {Huang}},
  \bibinfo {author} {\bibfnamefont {S.-G.}\ \bibnamefont {Guan}}, \bibinfo
  {author} {\bibfnamefont {Y.-C.}\ \bibnamefont {Lai}}, \ and\ \bibinfo
  {author} {\bibfnamefont {C.~H.}\ \bibnamefont {Lai}},\ }\href@noop {}
  {\bibfield  {journal} {\bibinfo  {journal} {Chaos}\ }\textbf {\bibinfo
  {volume} {18}},\ \bibinfo {pages} {037117} (\bibinfo {year}
  {2008})}\BibitemShut {NoStop}%
\bibitem [{\citenamefont {Pecora}\ \emph {et~al.}(2014)\citenamefont {Pecora},
  \citenamefont {Sorrentino}, \citenamefont {Hagerstrom}, \citenamefont
  {Murphy},\ and\ \citenamefont {Roy}}]{PSHMR:2014}%
  \BibitemOpen
  \bibfield  {author} {\bibinfo {author} {\bibfnamefont {L.~M.}\ \bibnamefont
  {Pecora}}, \bibinfo {author} {\bibfnamefont {F.}~\bibnamefont {Sorrentino}},
  \bibinfo {author} {\bibfnamefont {A.~M.}\ \bibnamefont {Hagerstrom}},
  \bibinfo {author} {\bibfnamefont {T.~E.}\ \bibnamefont {Murphy}}, \ and\
  \bibinfo {author} {\bibfnamefont {R.}~\bibnamefont {Roy}},\ }\href@noop {}
  {\bibfield  {journal} {\bibinfo  {journal} {Nature Comm.}\ }\textbf {\bibinfo
  {volume} {5}},\ \bibinfo {pages} {4079} (\bibinfo {year} {2014})}\BibitemShut
  {NoStop}%
\bibitem [{\citenamefont {Dorogovtsev}\ and\ \citenamefont
  {Mendes}(2002)}]{DM:2002}%
  \BibitemOpen
  \bibfield  {author} {\bibinfo {author} {\bibfnamefont {S.~N.}\ \bibnamefont
  {Dorogovtsev}}\ and\ \bibinfo {author} {\bibfnamefont {J.~F.~F.}\
  \bibnamefont {Mendes}},\ }\href@noop {} {\bibfield  {journal} {\bibinfo
  {journal} {Adv. Phys.}\ }\textbf {\bibinfo {volume} {51}},\ \bibinfo {pages}
  {1079} (\bibinfo {year} {2002})}\BibitemShut {NoStop}%
\bibitem [{\citenamefont {Holme}\ and\ \citenamefont {Newman}(2006)}]{HN:2006}%
  \BibitemOpen
  \bibfield  {author} {\bibinfo {author} {\bibfnamefont {P.}~\bibnamefont
  {Holme}}\ and\ \bibinfo {author} {\bibfnamefont {M.~E.~J.}\ \bibnamefont
  {Newman}},\ }\href {\doibase 10.1103/PhysRevE.74.056108} {\bibfield
  {journal} {\bibinfo  {journal} {Phys. Rev. E}\ }\textbf {\bibinfo {volume}
  {74}},\ \bibinfo {pages} {056108} (\bibinfo {year} {2006})}\BibitemShut
  {NoStop}%
\bibitem [{\citenamefont {Stilwell}\ \emph {et~al.}(2006)\citenamefont
  {Stilwell}, \citenamefont {Bollt},\ and\ \citenamefont
  {Roberson}}]{SBR:2006}%
  \BibitemOpen
  \bibfield  {author} {\bibinfo {author} {\bibfnamefont {D.~J.}\ \bibnamefont
  {Stilwell}}, \bibinfo {author} {\bibfnamefont {E.~M.}\ \bibnamefont {Bollt}},
  \ and\ \bibinfo {author} {\bibfnamefont {D.~G.}\ \bibnamefont {Roberson}},\
  }\href@noop {} {\bibfield  {journal} {\bibinfo  {journal} {SIAM J. Appl. Dyn.
  Syst.}\ }\textbf {\bibinfo {volume} {5}},\ \bibinfo {pages} {347} (\bibinfo
  {year} {2006})}\BibitemShut {NoStop}%
\bibitem [{\citenamefont {Porfiri}\ \emph {et~al.}(2006)\citenamefont
  {Porfiri}, \citenamefont {Stilwell}, \citenamefont {Bollt},\ and\
  \citenamefont {Skufca}}]{PSBS:2006}%
  \BibitemOpen
  \bibfield  {author} {\bibinfo {author} {\bibfnamefont {M.}~\bibnamefont
  {Porfiri}}, \bibinfo {author} {\bibfnamefont {D.~J.}\ \bibnamefont
  {Stilwell}}, \bibinfo {author} {\bibfnamefont {E.~M.}\ \bibnamefont {Bollt}},
  \ and\ \bibinfo {author} {\bibfnamefont {J.~D.}\ \bibnamefont {Skufca}},\
  }\href@noop {} {\bibfield  {journal} {\bibinfo  {journal} {Physica D}\
  }\textbf {\bibinfo {volume} {224}},\ \bibinfo {pages} {102} (\bibinfo {year}
  {2006})}\BibitemShut {NoStop}%
\bibitem [{\citenamefont {Kim}\ \emph {et~al.}(2013)\citenamefont {Kim},
  \citenamefont {Do},\ and\ \citenamefont {Lai}}]{KDL:2013}%
  \BibitemOpen
  \bibfield  {author} {\bibinfo {author} {\bibfnamefont {B.}~\bibnamefont
  {Kim}}, \bibinfo {author} {\bibfnamefont {Y.}~\bibnamefont {Do}}, \ and\
  \bibinfo {author} {\bibfnamefont {Y.-C.}\ \bibnamefont {Lai}},\ }\href
  {\doibase 10.1103/PhysRevE.88.042818} {\bibfield  {journal} {\bibinfo
  {journal} {Phys. Rev. E}\ }\textbf {\bibinfo {volume} {88}},\ \bibinfo
  {pages} {042818} (\bibinfo {year} {2013})}\BibitemShut {NoStop}%
\bibitem [{\citenamefont {Gleiser}\ and\ \citenamefont
  {Zanette}(2006)}]{GZ:2006}%
  \BibitemOpen
  \bibfield  {author} {\bibinfo {author} {\bibfnamefont {P.~M.}\ \bibnamefont
  {Gleiser}}\ and\ \bibinfo {author} {\bibfnamefont {D.~H.}\ \bibnamefont
  {Zanette}},\ }\href@noop {} {\bibfield  {journal} {\bibinfo  {journal} {Eur.
  Phys. J. B}\ }\textbf {\bibinfo {volume} {53}},\ \bibinfo {pages} {233}
  (\bibinfo {year} {2006})}\BibitemShut {NoStop}%
\bibitem [{\citenamefont {Robinson}\ \emph {et~al.}(2009)\citenamefont
  {Robinson}, \citenamefont {Henderson}, \citenamefont {Matar}, \citenamefont
  {Riley},\ and\ \citenamefont {Gray}}]{RHMRG:2009}%
  \BibitemOpen
  \bibfield  {author} {\bibinfo {author} {\bibfnamefont {P.~A.}\ \bibnamefont
  {Robinson}}, \bibinfo {author} {\bibfnamefont {J.~A.}\ \bibnamefont
  {Henderson}}, \bibinfo {author} {\bibfnamefont {E.}~\bibnamefont {Matar}},
  \bibinfo {author} {\bibfnamefont {P.}~\bibnamefont {Riley}}, \ and\ \bibinfo
  {author} {\bibfnamefont {R.~T.}\ \bibnamefont {Gray}},\ }\href {\doibase
  10.1103/PhysRevLett.103.108104} {\bibfield  {journal} {\bibinfo  {journal}
  {Phys. Rev. Lett.}\ }\textbf {\bibinfo {volume} {103}},\ \bibinfo {pages}
  {108104} (\bibinfo {year} {2009})}\BibitemShut {NoStop}%
\bibitem [{\citenamefont {Pan}\ and\ \citenamefont {Sinha}(2007)}]{KS:2007}%
  \BibitemOpen
  \bibfield  {author} {\bibinfo {author} {\bibfnamefont {R.~K.}\ \bibnamefont
  {Pan}}\ and\ \bibinfo {author} {\bibfnamefont {S.}~\bibnamefont {Sinha}},\
  }\href {\doibase 10.1103/PhysRevE.76.045103} {\bibfield  {journal} {\bibinfo
  {journal} {Phys. Rev. E}\ }\textbf {\bibinfo {volume} {76}},\ \bibinfo
  {pages} {045103} (\bibinfo {year} {2007})}\BibitemShut {NoStop}%
\bibitem [{\citenamefont {Sorrentino}\ and\ \citenamefont
  {Ott}(2008)}]{SO:2008}%
  \BibitemOpen
  \bibfield  {author} {\bibinfo {author} {\bibfnamefont {F.}~\bibnamefont
  {Sorrentino}}\ and\ \bibinfo {author} {\bibfnamefont {E.}~\bibnamefont
  {Ott}},\ }\href {\doibase 10.1103/PhysRevLett.100.114101} {\bibfield
  {journal} {\bibinfo  {journal} {Phys. Rev. Lett.}\ }\textbf {\bibinfo
  {volume} {100}},\ \bibinfo {pages} {114101} (\bibinfo {year}
  {2008})}\BibitemShut {NoStop}%
\bibitem [{\citenamefont {Zhou}\ and\ \citenamefont {Kurths}(2006)}]{ZK:2006}%
  \BibitemOpen
  \bibfield  {author} {\bibinfo {author} {\bibfnamefont {C.}~\bibnamefont
  {Zhou}}\ and\ \bibinfo {author} {\bibfnamefont {J.}~\bibnamefont {Kurths}},\
  }\href {\doibase 10.1103/PhysRevLett.96.164102} {\bibfield  {journal}
  {\bibinfo  {journal} {Phys. Rev. Lett.}\ }\textbf {\bibinfo {volume} {96}},\
  \bibinfo {pages} {164102} (\bibinfo {year} {2006})}\BibitemShut {NoStop}%
\bibitem [{\citenamefont {Li}\ \emph {et~al.}(2011{\natexlab{a}})\citenamefont
  {Li}, \citenamefont {Guan},\ and\ \citenamefont {Lai}}]{LGL:2011}%
  \BibitemOpen
  \bibfield  {author} {\bibinfo {author} {\bibfnamefont {M.}~\bibnamefont
  {Li}}, \bibinfo {author} {\bibfnamefont {S.-G.}\ \bibnamefont {Guan}}, \ and\
  \bibinfo {author} {\bibfnamefont {C.-H.}\ \bibnamefont {Lai}},\ }\href@noop
  {} {\bibfield  {journal} {\bibinfo  {journal} {EPL}\ }\textbf {\bibinfo
  {volume} {96}},\ \bibinfo {pages} {58004} (\bibinfo {year}
  {2011}{\natexlab{a}})}\BibitemShut {NoStop}%
\bibitem [{\citenamefont {Son}\ \emph {et~al.}(2009)\citenamefont {Son},
  \citenamefont {Kim}, \citenamefont {Hong},\ and\ \citenamefont
  {Jeong}}]{SKHJ:2009}%
  \BibitemOpen
  \bibfield  {author} {\bibinfo {author} {\bibfnamefont {S.-W.}\ \bibnamefont
  {Son}}, \bibinfo {author} {\bibfnamefont {B.~J.}\ \bibnamefont {Kim}},
  \bibinfo {author} {\bibfnamefont {H.}~\bibnamefont {Hong}}, \ and\ \bibinfo
  {author} {\bibfnamefont {H.}~\bibnamefont {Jeong}},\ }\href {\doibase
  10.1103/PhysRevLett.103.228702} {\bibfield  {journal} {\bibinfo  {journal}
  {Phys. Rev. Lett.}\ }\textbf {\bibinfo {volume} {103}},\ \bibinfo {pages}
  {228702} (\bibinfo {year} {2009})}\BibitemShut {NoStop}%
\bibitem [{\citenamefont {Li}\ \emph {et~al.}(2011{\natexlab{b}})\citenamefont
  {Li}, \citenamefont {Wang}, \citenamefont {Fan}, \citenamefont {Di},\ and\
  \citenamefont {Lai}}]{LWFDL:2011}%
  \BibitemOpen
  \bibfield  {author} {\bibinfo {author} {\bibfnamefont {M.}~\bibnamefont
  {Li}}, \bibinfo {author} {\bibfnamefont {X.-G.}\ \bibnamefont {Wang}},
  \bibinfo {author} {\bibfnamefont {Y.}~\bibnamefont {Fan}}, \bibinfo {author}
  {\bibfnamefont {Z.}~\bibnamefont {Di}}, \ and\ \bibinfo {author}
  {\bibfnamefont {C.-H.}\ \bibnamefont {Lai}},\ }\href@noop {} {\bibfield
  {journal} {\bibinfo  {journal} {Chaos}\ }\textbf {\bibinfo {volume} {21}},\
  \bibinfo {pages} {025108} (\bibinfo {year} {2011}{\natexlab{b}})}\BibitemShut
  {NoStop}%
\bibitem [{\citenamefont {Butts}(2009)}]{Butts:2009}%
  \BibitemOpen
  \bibfield  {author} {\bibinfo {author} {\bibfnamefont {C.~T.}\ \bibnamefont
  {Butts}},\ }\href@noop {} {\bibfield  {journal} {\bibinfo  {journal}
  {Science}\ }\textbf {\bibinfo {volume} {325}},\ \bibinfo {pages} {414}
  (\bibinfo {year} {2009})}\BibitemShut {NoStop}%
\bibitem [{\citenamefont {Motter}\ \emph {et~al.}(2005)\citenamefont {Motter},
  \citenamefont {Zhou},\ and\ \citenamefont {Kurths}}]{MZK:2005}%
  \BibitemOpen
  \bibfield  {author} {\bibinfo {author} {\bibfnamefont {A.~E.}\ \bibnamefont
  {Motter}}, \bibinfo {author} {\bibfnamefont {C.~S.}\ \bibnamefont {Zhou}}, \
  and\ \bibinfo {author} {\bibfnamefont {J.}~\bibnamefont {Kurths}},\
  }\href@noop {} {\bibfield  {journal} {\bibinfo  {journal} {Europhys. Lett.}\
  }\textbf {\bibinfo {volume} {69}},\ \bibinfo {pages} {334} (\bibinfo {year}
  {2005})}\BibitemShut {NoStop}%
\bibitem [{\citenamefont {Wang}\ \emph
  {et~al.}(2007{\natexlab{b}})\citenamefont {Wang}, \citenamefont {Lai},\ and\
  \citenamefont {Lai}}]{WLL:2007}%
  \BibitemOpen
  \bibfield  {author} {\bibinfo {author} {\bibfnamefont {X.-G.}\ \bibnamefont
  {Wang}}, \bibinfo {author} {\bibfnamefont {Y.-C.}\ \bibnamefont {Lai}}, \
  and\ \bibinfo {author} {\bibfnamefont {C.-H.}\ \bibnamefont {Lai}},\
  }\href@noop {} {\bibfield  {journal} {\bibinfo  {journal} {Phys. Rev. E}\
  }\textbf {\bibinfo {volume} {75}},\ \bibinfo {pages} {056205} (\bibinfo
  {year} {2007}{\natexlab{b}})}\BibitemShut {NoStop}%
\bibitem [{\citenamefont {Pecora}\ and\ \citenamefont
  {Carroll}(2015)}]{PC:2015}%
  \BibitemOpen
  \bibfield  {author} {\bibinfo {author} {\bibfnamefont {L.~M.}\ \bibnamefont
  {Pecora}}\ and\ \bibinfo {author} {\bibfnamefont {T.}~\bibnamefont
  {Carroll}},\ }\href@noop {} {\bibfield  {journal} {\bibinfo  {journal}
  {Chaos}\ }\textbf {\bibinfo {volume} {25}},\ \bibinfo {pages} {097611}
  (\bibinfo {year} {2015})}\BibitemShut {NoStop}%
\bibitem [{\citenamefont {Bak}\ \emph {et~al.}(1987)\citenamefont {Bak},
  \citenamefont {Tang},\ and\ \citenamefont {Wiesenfeld}}]{BTW:1987}%
  \BibitemOpen
  \bibfield  {author} {\bibinfo {author} {\bibfnamefont {P.}~\bibnamefont
  {Bak}}, \bibinfo {author} {\bibfnamefont {C.}~\bibnamefont {Tang}}, \ and\
  \bibinfo {author} {\bibfnamefont {K.}~\bibnamefont {Wiesenfeld}},\ }\href
  {\doibase 10.1103/PhysRevLett.59.381} {\bibfield  {journal} {\bibinfo
  {journal} {Phys. Rev. Lett.}\ }\textbf {\bibinfo {volume} {59}},\ \bibinfo
  {pages} {381} (\bibinfo {year} {1987})}\BibitemShut {NoStop}%
\bibitem [{\citenamefont {Stanley}(1987)}]{Stanley:book}%
  \BibitemOpen
  \bibfield  {author} {\bibinfo {author} {\bibfnamefont {H.~E.}\ \bibnamefont
  {Stanley}},\ }\href@noop {} {\emph {\bibinfo {title} {Introduction to Phase
  Transitions and Critical Phenomena}}}\ (\bibinfo  {publisher} {Oxford
  University Press},\ \bibinfo {address} {Oxford, UK},\ \bibinfo {year}
  {1987})\BibitemShut {NoStop}%
\bibitem [{\citenamefont {Motter}\ \emph {et~al.}(2013)\citenamefont {Motter},
  \citenamefont {Myers}, \citenamefont {Anghel},\ and\ \citenamefont
  {Nishikawa}}]{MMAN:2013}%
  \BibitemOpen
  \bibfield  {author} {\bibinfo {author} {\bibfnamefont {A.~E.}\ \bibnamefont
  {Motter}}, \bibinfo {author} {\bibfnamefont {S.~A.}\ \bibnamefont {Myers}},
  \bibinfo {author} {\bibfnamefont {M.}~\bibnamefont {Anghel}}, \ and\ \bibinfo
  {author} {\bibfnamefont {T.}~\bibnamefont {Nishikawa}},\ }\href@noop {}
  {\bibfield  {journal} {\bibinfo  {journal} {Nature Phys.}\ }\textbf {\bibinfo
  {volume} {9}},\ \bibinfo {pages} {191} (\bibinfo {year} {2013})}\BibitemShut
  {NoStop}%
\bibitem [{\citenamefont {Pagani}\ and\ \citenamefont
  {Aiello}(2013)}]{PA:2013}%
  \BibitemOpen
  \bibfield  {author} {\bibinfo {author} {\bibfnamefont {G.~A.}\ \bibnamefont
  {Pagani}}\ and\ \bibinfo {author} {\bibfnamefont {M.}~\bibnamefont
  {Aiello}},\ }\href@noop {} {\bibfield  {journal} {\bibinfo  {journal}
  {Physica A}\ }\textbf {\bibinfo {volume} {392}},\ \bibinfo {pages} {2699}
  (\bibinfo {year} {2013})}\BibitemShut {NoStop}%
\bibitem [{\citenamefont {Pecora}\ and\ \citenamefont
  {Carroll}(1998)}]{PC:1998}%
  \BibitemOpen
  \bibfield  {author} {\bibinfo {author} {\bibfnamefont {L.~M.}\ \bibnamefont
  {Pecora}}\ and\ \bibinfo {author} {\bibfnamefont {T.~L.}\ \bibnamefont
  {Carroll}},\ }\href@noop {} {\bibfield  {journal} {\bibinfo  {journal} {Phys.
  Rev. Lett.}\ }\textbf {\bibinfo {volume} {80}},\ \bibinfo {pages} {2109}
  (\bibinfo {year} {1998})}\BibitemShut {NoStop}%
\bibitem [{\citenamefont {Hu}\ \emph {et~al.}(1998)\citenamefont {Hu},
  \citenamefont {Yang},\ and\ \citenamefont {Liu}}]{HYL:1998}%
  \BibitemOpen
  \bibfield  {author} {\bibinfo {author} {\bibfnamefont {G.}~\bibnamefont
  {Hu}}, \bibinfo {author} {\bibfnamefont {J.}~\bibnamefont {Yang}}, \ and\
  \bibinfo {author} {\bibfnamefont {W.}~\bibnamefont {Liu}},\ }\href@noop {}
  {\bibfield  {journal} {\bibinfo  {journal} {Phys. Rev. E}\ }\textbf {\bibinfo
  {volume} {58}},\ \bibinfo {pages} {4440} (\bibinfo {year}
  {1998})}\BibitemShut {NoStop}%
\bibitem [{\citenamefont {Huang}\ \emph {et~al.}(2009)\citenamefont {Huang},
  \citenamefont {Chen}, \citenamefont {Lai},\ and\ \citenamefont
  {Pecora}}]{HCLP:2009}%
  \BibitemOpen
  \bibfield  {author} {\bibinfo {author} {\bibfnamefont {L.}~\bibnamefont
  {Huang}}, \bibinfo {author} {\bibfnamefont {Q.-F.}\ \bibnamefont {Chen}},
  \bibinfo {author} {\bibfnamefont {Y.-C.}\ \bibnamefont {Lai}}, \ and\
  \bibinfo {author} {\bibfnamefont {L.~M.}\ \bibnamefont {Pecora}},\
  }\href@noop {} {\bibfield  {journal} {\bibinfo  {journal} {Phys. Rev. E}\
  }\textbf {\bibinfo {volume} {80}},\ \bibinfo {pages} {036204} (\bibinfo
  {year} {2009})}\BibitemShut {NoStop}%
\bibitem [{\citenamefont {Wackerbauer}(2007)}]{Wackerbauer:2007}%
  \BibitemOpen
  \bibfield  {author} {\bibinfo {author} {\bibfnamefont {R.}~\bibnamefont
  {Wackerbauer}},\ }\href {\doibase 10.1103/PhysRevE.76.056207} {\bibfield
  {journal} {\bibinfo  {journal} {Phys. Rev. E}\ }\textbf {\bibinfo {volume}
  {76}},\ \bibinfo {pages} {056207} (\bibinfo {year} {2007})}\BibitemShut
  {NoStop}%
\bibitem [{\citenamefont {Qi}\ \emph {et~al.}(2008)\citenamefont {Qi},
  \citenamefont {Huang}, \citenamefont {Shen}, \citenamefont {Wang},\ and\
  \citenamefont {Chen}}]{QHSWC:2008}%
  \BibitemOpen
  \bibfield  {author} {\bibinfo {author} {\bibfnamefont {G.~X.}\ \bibnamefont
  {Qi}}, \bibinfo {author} {\bibfnamefont {H.~B.}\ \bibnamefont {Huang}},
  \bibinfo {author} {\bibfnamefont {C.~K.}\ \bibnamefont {Shen}}, \bibinfo
  {author} {\bibfnamefont {H.~J.}\ \bibnamefont {Wang}}, \ and\ \bibinfo
  {author} {\bibfnamefont {L.}~\bibnamefont {Chen}},\ }\href {\doibase
  10.1103/PhysRevE.77.056205} {\bibfield  {journal} {\bibinfo  {journal} {Phys.
  Rev. E}\ }\textbf {\bibinfo {volume} {77}},\ \bibinfo {pages} {056205}
  (\bibinfo {year} {2008})}\BibitemShut {NoStop}%
\bibitem [{\citenamefont {Brabow}\ \emph {et~al.}(2011)\citenamefont {Brabow},
  \citenamefont {Grosskinsky}, ,\ and\ \citenamefont {Timme}}]{BGT:2008}%
  \BibitemOpen
  \bibfield  {author} {\bibinfo {author} {\bibfnamefont {C.}~\bibnamefont
  {Brabow}}, \bibinfo {author} {\bibfnamefont {S.}~\bibnamefont {Grosskinsky}},
  , \ and\ \bibinfo {author} {\bibfnamefont {M.}~\bibnamefont {Timme}},\
  }\href@noop {} {\bibfield  {journal} {\bibinfo  {journal} {Eur. Phys. J. B}\
  }\textbf {\bibinfo {volume} {84}},\ \bibinfo {pages} {613} (\bibinfo {year}
  {2011})}\BibitemShut {NoStop}%
\bibitem [{\citenamefont {Fu}\ \emph {et~al.}(2012)\citenamefont {Fu},
  \citenamefont {Zhang}, \citenamefont {Zhan},\ and\ \citenamefont
  {Wang}}]{FZZW:2012}%
  \BibitemOpen
  \bibfield  {author} {\bibinfo {author} {\bibfnamefont {C.}~\bibnamefont
  {Fu}}, \bibinfo {author} {\bibfnamefont {H.}~\bibnamefont {Zhang}}, \bibinfo
  {author} {\bibfnamefont {M.}~\bibnamefont {Zhan}}, \ and\ \bibinfo {author}
  {\bibfnamefont {X.}~\bibnamefont {Wang}},\ }\href {\doibase
  10.1103/PhysRevE.85.066208} {\bibfield  {journal} {\bibinfo  {journal} {Phys.
  Rev. E}\ }\textbf {\bibinfo {volume} {85}},\ \bibinfo {pages} {066208}
  (\bibinfo {year} {2012})}\BibitemShut {NoStop}%
\bibitem [{\citenamefont {Barab\'{a}si}\ \emph {et~al.}(1999)\citenamefont
  {Barab\'{a}si}, \citenamefont {Albert},\ and\ \citenamefont
  {Jeong}}]{BAJ:1999}%
  \BibitemOpen
  \bibfield  {author} {\bibinfo {author} {\bibfnamefont {A.-L.}\ \bibnamefont
  {Barab\'{a}si}}, \bibinfo {author} {\bibfnamefont {R.}~\bibnamefont
  {Albert}}, \ and\ \bibinfo {author} {\bibfnamefont {H.}~\bibnamefont
  {Jeong}},\ }\href@noop {} {\bibfield  {journal} {\bibinfo  {journal} {Physica
  A}\ }\textbf {\bibinfo {volume} {272}},\ \bibinfo {pages} {173} (\bibinfo
  {year} {1999})}\BibitemShut {NoStop}%
\bibitem [{\citenamefont {R\"{o}ssler}(1976)}]{Rossler:1976}%
  \BibitemOpen
  \bibfield  {author} {\bibinfo {author} {\bibfnamefont {O.~E.}\ \bibnamefont
  {R\"{o}ssler}},\ }\href@noop {} {\bibfield  {journal} {\bibinfo  {journal}
  {Phys. Lett. A}\ }\textbf {\bibinfo {volume} {57}},\ \bibinfo {pages} {397}
  (\bibinfo {year} {1976})}\BibitemShut {NoStop}%
\end{thebibliography}%

\end{document}